\documentclass[10pt,a4paper]{article}

\usepackage{pdflscape}

\usepackage[T1]{fontenc}
\usepackage[utf8]{inputenc}
\usepackage{graphicx}
\usepackage{multicol}
\usepackage{tabto}
\usepackage{url}
\usepackage{enumitem}
\usepackage{textcomp}
\usepackage{eurosym}
\usepackage[labelfont=bf]{caption}
\usepackage[yyyymmdd]{datetime}
\usepackage[strings]{underscore}
\usepackage{todonotes}
\usepackage{amsmath}
\usepackage{amssymb}

\usepackage{eso-pic}
\usepackage{microtype}
\usepackage[scaled=0.85]{beramono}

\usepackage{tikz}
\usepackage{flowchart}
\usetikzlibrary{shapes,arrows.meta,chains,external,positioning,calc}

\usepackage{listings}

\lstdefinestyle{compactrust}{
    basicstyle=\ttfamily\fontsize{6.4}{6.8}\selectfont,
    breaklines=true,
    frame=single,
    columns=fullflexible,
    keepspaces=true,
    showstringspaces=false,
    aboveskip=0pt,
    belowskip=0pt,
    lineskip=-1pt
}

\usepackage{float}

\newcommand{\eat}[1]{}

\AddToShipoutPictureFG{%
  \AtPageLowerLeft{%
    \hspace*{5mm}%
    \raisebox{0.5\paperheight}[0pt][0pt]{%
      \rotatebox[origin=c]{90}{%
        \ttfamily
        \fontsize{7}{8}\selectfont
        \textls[120]{WORKING PAPER \textbar\ UPDATED \MakeUppercase{\today}}%
      }%
    }%
  }%
}

\begin{document}

\bibliographystyle{unsrt}
\thispagestyle{empty}

\NumTabs{6}

\begin{center}
\huge{ Protocol-Embedded Compliance for Privacy-Preserving, Non-Custodial Digital Payments}\\
\end{center}

\begin{center}
\begin{minipage}[t]{0.32\linewidth}
\centering
\large{\bf Santiago De Simone}\\
Department of Computer Science\\
University College London\\
{\small \texttt{santiago.simone.24@alumni.ucl.ac.uk}}
\end{minipage}
\hfill
\begin{minipage}[t]{0.32\linewidth}
\centering
\large{\bf Geoffrey Goodell}\\
Department of Computer
Science\\
University College London\\
{\small \texttt{g.goodell@ucl.ac.uk} }
\end{minipage}
\hfill
\begin{minipage}[t]{0.32\linewidth}
\centering
\large{\bf Georgios Samakovitis}\\
School of Computing \& Mathematical Sciences\\
University of Greenwich\\
{\small \texttt{g.samakovitis@greenwich.ac.uk}}
\end{minipage}
\end{center}

\begin{center}
\begin{minipage}{0.80\linewidth}

\textbf{Abstract---}Received wisdom on payments infrastructure strongly supports the custodial, account-based model as a necessity for transaction integrity, auditability and verification; the set of fundamental primitives for regulated digital money exchange, the argument goes, necessitates designated identifiable entities that store and process credentials, perform KYC, and ultimately act as the 'single version of the truth' for compliance remediation and, most important, AML. In this paper, we propose this is not the case, by arguing that non-custodial, cash-like digital assets can embody such capabilities, in an arguably more secure manner. 
\par\smallskip

We present a reference architecture and core protocol rules for digital-value-exchange
systems that preserve meaningful user privacy while enabling strong auditability, and we
show that this can be made compatible with existing compliance practice. The protocol
defines the conditions under which digital asset creation, transfer, and redemption are
valid; the architecture specifies the allocation of actors, roles and components through
which these rules operate, enabling independent verification of transaction compliance
with applicable norms. Building upon the Unforgeable, Stateful, Oblivious (USO) asset
model of Goodell et al., assets are held by their users rather than represented on a
ledger; the ledger records only commitments to asset state as it evolves, and simple
oblivious notaries validate each state transition. We embed regulatory compliance data
directly into the asset state as cryptographically signed attestations issued by
appropriately authorized entities. A transfer is valid only if the applicable compliance
predicates are satisfied and the resulting signature is included in the asset state. Compliance is enforced at the protocol level rather than through institutional custody
or identity-based account control. We conclude that our proposed model can successfully interface with existing payment systems, making it possible to integrate non-custodial, compliance-verified transactions with legacy financial infrastructure.

\end{minipage}
\end{center}

\section{Introduction}

Prevailing approaches to regulated digital value exchange assume that auditability and AML
compliance require custodial intermediation or the accumulation of records by some
party~\cite{fatf_recommendations_2026,fatf_digital_identity_2020}. Such arrangements,
however, concentrate sensitive data in a single point of compromise, creating an
attractive target for attack and putting user privacy at risk.

This paper presents a reference architecture and core protocol rules for digital value
exchange systems that offer meaningful user privacy and strong auditability. Beyond the
design itself, the paper shows that non-custodial asset transfer can satisfy existing
compliance requirements. Its central contribution is the combination of three elements.
The first is a specification of the USO asset model, under which users hold assets that
carry their own state; a ledger records only commitments to that state, and simple
oblivious notaries validate state transitions, which preserves user privacy. The second is
verifiable attestations, cryptographically signed by appropriately authorized entities and
checkable by third parties, to support auditability. The third is the \textit{embedded
compliance} approach, under which compliance requirements are incorporated into asset
state transitions and evidenced through attestations carried within the asset state.
Compliance thus becomes part of the protocol-level conditions governing transfer: a valid
transfer requires that the applicable compliance predicates are satisfied and that
verifiable evidence of their satisfaction is produced.  Auditability and compliance
therefore rest on verifiable attestations held at the asset-state level, rather than on
custodial intermediation or accumulated records.

The remainder of the paper is organized as follows.  Section~\ref{sec:motivation} sets out
the motivation, and Section~\ref{sec:background-and-related-work} reviews background and
related work. Section~\ref{sec:reference-architecture} presents our proposed solution,
covering its design objectives and the actors, components, and interfaces through which
the protocol is operationalized.  Section~\ref{sec:core-protocol-rules} formalizes the
protocol rules governing asset creation, transfer, and redemption.
Section~\ref{sec:use-cases} presents use cases, including retail payments and more general
digital asset exchange, to show how the solution operates in realistic environments.
Section~\ref{sec:discussion} discusses implications, and Section~\ref{sec:conclusion}
concludes with directions for future work.

\section{Motivation}
\label{sec:motivation}

A new generation of digital-value-exchange systems (or simply payment systems) must ideally reconcile two seemingly conflicting objectives: protecting user privacy and ensuring effective auditability, meaning that these systems can demonstrably verify that transactions conform to regulatory obligations.

The traditional payments paradigm resolves this tension in favour of oversight.
Custodial, account-based arrangements give asset custodians a critical role in
verifying payment integrity, and the custodian already has full knowledge of
user data by default; privacy is therefore a matter of policy and data
minimisation, not an architectural feature. Cryptographic solutions are
primarily introduced to narrow identity disclosure to the auditor and regulator,
not to limit what the custodian itself knows.

At the other extreme, fully anonymous instruments provide privacy but do not
permit effective enforcement of compliance obligations. Existing approaches
therefore trade privacy for compliance, or vice versa.

What drives this work is a set of unresolved questions at the intersection of digital payments, privacy, and regulatory compliance. Can a payment system offer meaningful user privacy without forfeiting regulatory enforceability? Is custodial oversight a technical necessity for auditability, or merely a historical artifact of custodial, account-based design? And can existing financial institutions continue to play a role in a system that no longer requires them to act as custodians for every transaction? These questions frame the problem space addressed in this article.

Many systems have tried to reconcile privacy and
compliance, but each family of approaches carries a cost: regulator back doors expose otherwise private payments; user-held state (e.g.,
balances, note histories) can later be compromised and linked to past
transactions; single-operator designs concentrate trust in one party that
issues, settles, and checks compliance; and ledger-based designs evaluate compliance predicates over a global state maintained on a ledger, requiring heavyweight infrastructure and, as with
user-held state, leaving a persistent record whose breach would expose users'
transactional data. Section~\ref{sec:background-and-related-work} describes these families in
detail and positions this work against each.

Furthermore, verifiable compliance for digital assets is necessarily subject to a complex set of requirements. Indicatively, de facto data models for originator and beneficiary information \cite{iso20022,ivms101,fatfRecommendation16}, as well as proposed weekly and quarterly supervisory reports for permitted payment stablecoin issuers \cite{fdic2026geniusReporting}, are standard features of present custodial payments infrastructure. They all involve identity disclosure of one form or another. More crucially, when interfacing with non-custodial, self-hosted wallets, compliance becomes problematic: verification reverts to cryptographic ownership proofs, such as use of a signed message or a micro-transaction ``Satoshi test'' to prove control of a key, rather than the identity behind it. That in itself implies a regulatory gap, since existing custodial models rely on intermediary identity-record knowledge, which is obscured during cryptographic ownership proofs. We propose that such restrictions may be obviated by introducing compliance as an infrastructure feature within transaction protocols.

We adopt a token-based design that renders payments cryptographically unlinkable to their payers, following the approach first described by Chaum~\cite{chaum_blind_signatures_1983}, which we believe is the best approach to ensuring that payers have privacy by design. We then address the challenge of decoupling compliance enforcement from custodial control. To do so, we base our token-level architectural design on the Unforgeable, Stateful, Oblivious (USO) asset model \cite{goodell_toliver_nakib_scalable_payments_2021} because it enables the embedding of regulatory-relevant data within the assets themselves, with the data persisting as the asset state evolves.

The USO model is also lightweight in its infrastructure requirements. Assets are held directly by their users, at the edge of the system, rather than being represented on a ledger, and validation requires only simple oblivious notaries; consequently, no single operator is required to issue assets, settle transactions, and verify compliance. The proposed architecture therefore assumes neither global state nor a distributed ledger technology system. Compliance evaluation accordingly takes place off-ledger: attestations are issued by recognised parties, and the protocol rules are verified by the parties that rely on an asset, rather than being executed by on-chain oracles or smart contracts. 

Building on this foundation, we describe a reference architecture and core protocol rules for compliant, token-based digital value transfers in which each proposed USO-like asset state transition is evaluated against a set of rules applicable to that transition. If the proposed state transition is conformant, then an attestation is produced and embedded within the asset's new state. To that end, we are utilising and extending the approach of embedded compliance, a concept first introduced in the context of decentralised programmable assets by Samakovitis~\cite{samakovitis2023decentralized,samakovitisEmbeddedCompliance}, where it is proposed that regulatory compliance be transformed from a verification process into an asset feature. Assets that do not carry evidence of compliance when mandatory are treated as incompatible by the protocol.

Our design elevates compliance verification from a discretionary function
performed by custodial intermediaries to a protocol-level mechanism that
operates independently of custody, thereby showing how auditability can be
achieved for non-custodial assets. In doing so, it replaces the prevailing
``forensic'' paradigm of compliance enforcement---where enforcement depends on
the \textit{ex post} reconstruction of transaction data---with a protocol-native
model in which compliance is enforced at execution time and cryptographically
bound to the asset itself.

The resulting solution interoperates with existing payment infrastructure,
redefines the roles of financial institutions and regulators, and supports
payment systems that combine strong compliance with meaningful user privacy and
self-custody, allowing lawful digital value transfers without requiring accounts
at either end of a transaction.

\section{Background and Related Work}
\label{sec:background-and-related-work}

Contemporary retail digital payments are predominantly account-based. In modern monetary systems, most money used by households and firms takes the form of commercial-bank deposits, while central-bank money available to the public remains principally physical cash \cite{mcleay_money_creation_2014,chaum_how_to_issue_2021}. Electronic payments therefore ordinarily operate by modifying claims recorded in institutional accounts: the payer's balance is debited, the payee's balance is credited, and settlement between participating institutions ultimately relies upon central-bank money \cite{mcleay_money_creation_2014}. Several prominent retail-CBDC proposals preserve this general structure, assigning regulated payment service providers responsibility for managing user access and processing payment information even where the underlying monetary claim is a direct liability of the central bank \cite{auer_technology_retail_cbdc_2020,ecb_digital_euro_report_2020}.

Compliance within this environment is correspondingly implemented through obligations imposed upon regulated institutions. FATF recommendations require financial institutions and other covered entities to conduct customer due diligence, maintain records, and report suspicious transactions, while related guidance treats digital identity and transaction monitoring as instruments for satisfying these duties \cite{fatf_recommendations_2026,fatf_digital_identity_2020}. Compliance therefore operates principally through institutional processes applied to identified customers and transaction records. Recent work has examined privacy-enhancing technologies and access-control mechanisms as means of limiting the disclosure or subsequent use of such information \cite{fatf_new_technologies_2021,fatf_data_pooling_2021}. These approaches nevertheless generally presuppose that payer identity or transaction data are generated and made available within the compliance infrastructure. Privacy consequently remains dependent upon institutional governance and the appropriate use of information that already exists. By contrast, privacy by design requires that the architecture itself prevent the creation or disclosure of information capable of linking payers to their transactions, rather than relying upon assurances concerning how such information will subsequently be used.

A separate line of research has pursued this objective by representing digital money as transferable tokens rather than balances maintained within persistent accounts. Chaum's blind-signature construction established that an issuer can certify a digital token without learning the token that will later be presented in payment, thereby preventing its issuance from being linked to its subsequent use \cite{chaum_blind_signatures_1983}. More recent proposals extend this principle to retail digital currency, combining token-based issuance with two-tier institutional arrangements, non-custodial wallets, and mechanisms for verifying asset integrity or lawful reissuance \cite{chaum_how_to_issue_2021,goodell_toliver_nakib_scalable_payments_2021,friolo_goodell_toliver_nakib_private_payments_zk_reissuance_2024,bowler_non_custodial_wallet_2024}. Although these designs differ in their ledger models, trust assumptions, and allocation of regulatory functions, they share the objective of allowing users to possess and control digital monetary assets without revealing to the counterparty the banking account relationship that enabled token acquisition in the first place. GNU Taler \cite{dold_taler_2019,burdges_taler_2016} is a practical Chaumian system of this kind.

Several lines of work already pursue privacy and compliance together, so the combination is not new. Senn et al.~\cite{senn_sok_2026} systematise 36 payment system designs with potential for CBDC and give a broader comparison; in this section we group only the closest systems, by the cost we believe they impose on users and on the system as a whole.  Our design builds on USO assets~\cite{goodell_toliver_nakib_scalable_payments_2021}, which users hold themselves with no shared ledger, and adds signed attestations, carried in the asset, that each transfer meets the compliance requirements. Under each of the four costs below we describe how our design differs.

\textbf{Back doors.} Privacy coins with view keys let a key holder see transactions that are otherwise hidden \cite{zcash_protocol_spec}. Accountable variants of Zerocash add policy enforcement and selective tracing \cite{garman_accountable_privacy_2016}, and revocable-anonymity e-cash lets a trusted party lift a payer's anonymity \cite{stadler_fair_blind_signatures_1995,camenisch_fair_payment_1996}. In each, privacy holds only until some party chooses to look. Our design contains no such capability: compliance is evaluated on facts about the transfer, and the protocol itself provides no means of unilaterally recovering the payer from a payment.

\textbf{Users must hold state.} Platypus \cite{platypus_2022}, PEReDi \cite{peredi_2022} and KAIME \cite{kaime_2024} are account-based, so users keep long-lived secrets and account state. The auditable token scheme of Androulaki et al.~\cite{androulaki_auditable_tokens_2020} lets auditors connect a payer's transactions, PRCash \cite{wust_privacy_preserving_payments_2019} has payers keep a regulator-issued certificate and proofs of prior spending to enforce holding limits, and Zether \cite{zether_2020} keeps encrypted balances in accounts. State information of this kind can be demanded later, linked to past transactions, and used to coerce its holder; Friolo et al.~\cite{friolo_goodell_toliver_nakib_private_payments_zk_reissuance_2024} compare several of these designs on this point. In our design, the asset's state travels with the asset to its next holder, each transfer uses a fresh one-time key, and a payer is not required to retain anything that links them to a past payment.

\textbf{One indispensable operator.} GNU Taler \cite{dold_taler_2019,burdges_taler_2016} keeps payers anonymous, but one exchange issues the coins, settles deposits and applies compliance, so merchants must hand it functions they would rather control, much as if someone else kept their books. UTT \cite{tomescu_utt_2022} gives decentralised ecash with accountable privacy, but implements its ledger, bank, auditor and registration authorities through threshold cryptography and a Byzantine fault tolerant service, and limits each payer to an anonymity budget. Designs of this kind concentrate issuance, settlement and compliance checking in one operator or one heavy infrastructure, which then carries the same regulatory burden as the compliance bodies and holds substantial power, introducing a tension with privacy by design and system resilience.  Unlike GNU Taler, USO assets~\cite{goodell_toliver_nakib_scalable_payments_2021} do not require the issuer to facilitate every transaction or to apply compliance checks itself. In our solution, issuance, settlement, and compliance functions are decoupled (Section~\ref{sec:reference-architecture}), and institutions act on assets only at entry and exit.

\textbf{Compliance on a ledger.} A fourth family places the compliance predicate inside a ledger that holds the global state of all assets. Zexe \cite{zexe_2020} lets users prove offline that a transaction satisfies application-defined predicates, with regulation-friendly private stablecoins among its applications; Kachina \cite{kachina_2021} gives a model for private smart contracts, and Midnight \cite{midnight_docs_2026} combines a public ledger with locally held private state and selective disclosure; Privacy Pools \cite{buterin_privacy_pools_2024} let a withdrawer prove membership in an association set that demonstrates compliance; and ERC-3643 \cite{erc3643_2021} enforces identity and compliance checks inside permissioned tokens.  These designs do embed verifiable compliance in state transitions, and we do not claim that they are incompatible with the aims of this paper. All of them depend on that ledger.  USO does not: assets are held by their users with no representation on a shared ledger.  Client-side validation \cite{todd_client_side_validation_2016} and Plasma Cash \cite{buterin_plasma_cash_2018} also leave users holding the data for their own coins, but those systems are designed to anchor such data in a public blockchain. Our design does for USO assets what those systems do for global state ledgers, by embedding into assets compliance predicates representing conditions on the state transition.  The satisfaction of those conditions is evidenced by a signed attestation from an independent service that is embedded in the asset's state chain, not by a proof or contract check against a ledger's global state.

The USO model places assets outside the ledger and uses oblivious integrity services
together with asset-carried proofs of provenance
\cite{goodell_toliver_nakib_scalable_payments_2021,lynham_sark_2025}. To our knowledge,
none of the designs above combine all of the following: (i) assets held by users and not
represented on a ledger; (ii) no back door and no user-held state that links payers to
past payments; (iii) compliance attestations signed in real time by independent services
and carried in the asset; (iv) a role for existing financial institutions only at
withdrawal and redemption, where value enters and leaves the non-custodial domain; and (v)
operational decoupling of issuance, transaction processing, and compliance functions. It
is this combination of elements, rather than any single element, that makes our proposal
innovative.

\section{Reference Architecture}
\label{sec:reference-architecture}

We understand a protocol to specify allowed behaviour by defining what actions are possible and under what conditions they are valid. A protocol is independent of any particular system realization and may therefore admit multiple architectures and designs that implement its rules. The constraints imposed by a protocol delimit the class of admissible systems and determine the space within which architectural variation is permitted.

The reference architecture, by contrast, specifies the structural arrangement of actors, components, and interfaces through which protocol rules are implemented. It determines how protocol-level validity conditions are operationalized and which actors are responsible for producing, verifying, or recording the information required by the protocol.

The specification of the reference architecture proceeds in four parts: the design objectives that govern architectural choices; an integrated overview of the architecture and its principal operational sequence; the actors and roles that perform distinct functions within the system; and the components through which protocol operations are implemented.

\subsection{Design Objectives}

\subsubsection{Privacy by Design}

Our architecture design follows from the premise that non-custodial payment architectures are a necessary condition for meaningful user privacy, as custodial control structurally entails disclosure and continuous oversight of user activity. Existing designs are custodial, account-based, or both, and each is a problem for privacy. Custody means that someone else controls the user's assets. Accounts, even where nobody else holds the funds, oblige users to keep state-transition information that can later be demanded and linked to their past transactions. We therefore reject pure account-based designs, whether centralized or decentralized, and adopt a design in which users hold assets directly and are not required to retain such linking information.

Under these constraints, a token-based architecture is required. We therefore adopt a token-based design employing a Chaumian mint model of issuance \cite{chaum_blind_signatures_1983}. This construction prevents persistent linkage between payer identity and transaction history, ensures unlinkability between withdrawal and circulation, and precludes the formation of centralized transaction logs capable of reconstructing user-level activity profiles.

\subsubsection{Decoupling of Functions}

Traditional payment system architectures exhibit a structural dependence on custodial intermediaries for the enforcement of compliance obligations. The institution holding user funds performs identity verification, transaction monitoring, freezing, and reporting. When custody is removed (i.e., users hold assets directly), that enforcement channel disappears.

In our proposed architecture, compliance enforcement is transferred to the protocol instead of being exercised as a by-product of institutional custody. Regulatory authorities or financial institutions within their purview are still capable of conducting audits by relying on regulatory-relevant information recorded both within the asset state and the supporting infrastructure, which together provide sufficient evidentiary material to verify whether a transaction is compliant with the applicable norms.

Relocating compliance enforcement away from custodial control requires a structural separation of responsibilities. The reference architecture therefore organizes core functions into independent domains—issuance, payment processing (or transfer validation), redemption, and compliance verification—such that none presupposes custodial control and none is structurally dependent on the others for its operation.

\subsubsection{Auditability via Verifiability}

To address the challenge of decoupling compliance enforcement from custodial control, we base our token-level architectural design on the USO asset model \cite{goodell_toliver_nakib_scalable_payments_2021}, which enables the embedding of regulatory-relevant data within the assets themselves, with the data persisting as the asset state evolves.

As a result, auditability is achieved through verifiability: any authorized auditor is able to verify the conformance of transactions with the applicable rules that govern them through a mechanism based exclusively on the verification of compliance certificates embedded in the asset state chain. Confirming the presence of such certificates wherever the applicable protocol enforcement policy requires them constitutes the proof of compliance the audit seeks to establish.

Whether and where certificates are required is determined by the protocol's enforcement policy, which may mandate their presence at the point of asset transfer, redemption, or both — a distinction elaborated in Section~\ref{sec:core-protocol-rules}. Where enforcement applies, absence of the required certificate is treated by the protocol as non-conformance with the norms governing the proposed state change, rendering it invalid.

\subsubsection{Interoperability with Legacy Financial Infrastructure}

Our proposed architecture is designed to interoperate with existing financial infrastructure, supporting compliant digital-value exchange systems that preserve user privacy while enabling self-custody.

In our design, licensed financial institutions are positioned as the institutional interface through which token withdrawal (acquisition into self-custody) and token redemption (conversion back into conventional monetary form) are effected, thereby connecting the non-custodial digital asset environment with established monetary systems.

Through this interface, value held at licensed financial institutions—whether as deposits or other eligible financial instruments—is exchanged for USO-like digital assets issued under privacy-preserving protocols administered by its corresponding issuance authority. The authorization of acquisition is effected \textit{ex ante} by the financial institution through regulated procedures, analogous to a withdrawal of physical currency. In this way, tokens are acquired through regulated financial institutions acting as enforcement nodes, ensuring that entry into the self-custodial digital asset environment occurs in compliance with applicable legal and regulatory requirements.

Conversely, tokens may be converted into conventional monetary form through the same institutional interface. In accordance with the protocol rules (see Section~\ref{sec:core-protocol-rules}), redemption may require verification of the asset’s full state history through compliance certificates issued by the regulatory authority or by entities explicitly authorized by it. This ensures that only compliant assets are permitted to re-enter the conventional monetary system. Where assets fail admissibility verification, the receiving financial institution may assume conditional custody pending further review, potentially restricting, transforming, or withholding the resulting value until compliance issues are resolved.

Such interoperability between the non-custodial asset domain and the existing financial infrastructure is essential for three structural reasons.

\paragraph{Practical Adoption.}
By relying on established banking relationships as the interface between the self-custodial environment and the traditional custodial banking system, the architecture avoids the need for parallel identity systems or additional custodial intermediaries. This lowers onboarding complexity and supports incremental integration within existing financial structures.

\paragraph{Monetary Continuity.}
Interoperability ensures seamless convertibility between self-custodial digital assets and established monetary forms. The self-custodial asset domain remains integrated with the broader monetary infrastructure that supports wages, deposits, credit creation, settlement, and taxation.

\paragraph{Regulatory Integration.}
Because token acquisition and redemption occur exclusively within licensed institutions, existing legal and supervisory frameworks remain enforceable. Compliance verification is performed within institutional processes through protocol-level policy rules and verification of asset admissibility under applicable regulatory conditions, preserving continuity with current regulatory arrangements.

\subsection{General Architectural Overview}
\label{sec:general-architectural-overview}

Figure~\ref{fig:general-reference-architecture} provides an integrated structural and event-sequence view of the proposed reference architecture. It identifies the principal functional domains, actors, and components through which the protocol is implemented, while also following a representative asset lifecycle from acquisition and self-custodial holding, through compliance evaluation, transfer, and validation, to possible redemption. Existing banking relationships provide the entry and exit points through which users acquire tokens from, and redeem tokens back into, conventional monetary forms.

The figure is intended as an architectural overview rather than as an exhaustive specification of every permitted interaction. It shows how the financial interface, the non-custodial transfer environment, and the compliance and supervisory functions relate to one another within a representative operational sequence. The depicted flows therefore provide a common frame of reference for the more detailed descriptions that follow.

\begin{landscape}
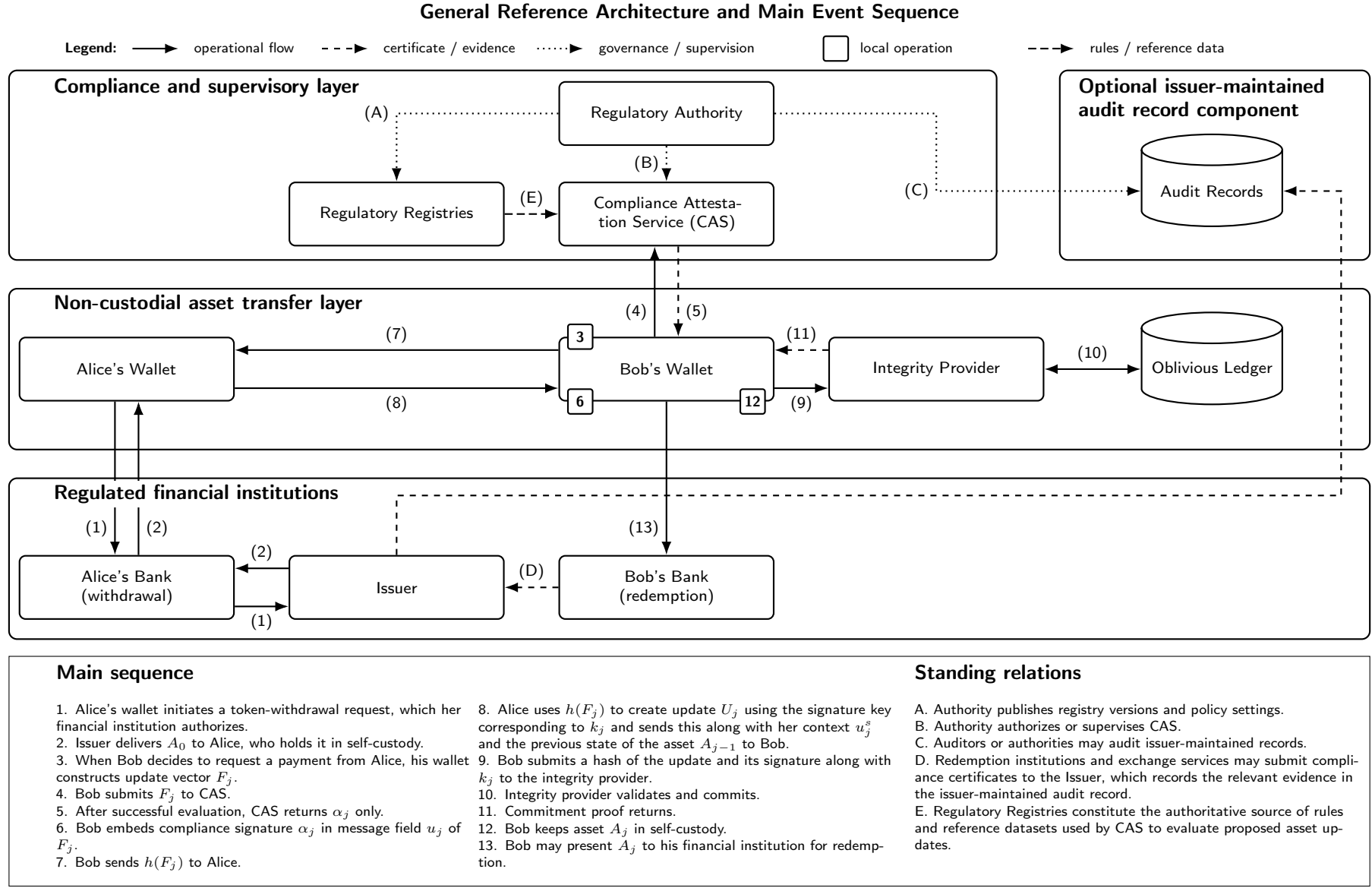
\begin{figure}[p]
\centering
\captionsetup{font=small}

\begin{tikzpicture}[
    x=1cm,
    y=1cm,
    font=\sffamily\footnotesize,
    >=Latex,
    actor/.style={
        draw,
        rounded corners=3pt,
        thick,
        align=center,
        minimum height=1.1cm,
        text width=3.5cm
    },
    component/.style={
        draw,
        thick,
        align=center,
        minimum height=0.62cm,
        text width=2.75cm,
        inner sep=2pt
    },
    store/.style={
        cylinder,
        shape border rotate=90,
        draw,
        thick,
        align=center,
        shape aspect=0.24,
        minimum height=1.6cm,
        minimum width=2.45cm,
        text width=2.10cm,
        inner sep=2pt
    },
    flow/.style={
        ->,
        thick
    },
    evidence/.style={
        ->,
        thick,
        dashed
    },
    governance/.style={
        ->,
        thick,
        dotted
    },
    reference/.style={
        ->,
        thick,
        dash pattern=on 4pt off 2pt
    },
    seqstep/.style={
        circle,
        draw,
        thick,
        fill=white,
        minimum size=4.1mm,
        inner sep=0pt,
        font=\sffamily\scriptsize\bfseries
    },
    localstep/.style={
        rectangle,
        rounded corners=1.2pt,
        draw,
        thick,
        fill=white,
        minimum size=4.3mm,
        inner sep=0pt,
        font=\sffamily\scriptsize\bfseries
    },
    eventtext/.style={
        anchor=north west,
        align=left,
        font=\sffamily\fontsize{7}{8.4}\selectfont
    }
]

% ------------------------------------------------------------------
% Title
% ------------------------------------------------------------------

\node[font=\sffamily\bfseries] at (12.1,13.2)
    {General Reference Architecture and Main Event Sequence};

% ------------------------------------------------------------------
% Legend
% ------------------------------------------------------------------

\node[anchor=west, font=\sffamily\scriptsize\bfseries] at (1.1,12.58)
    {Legend:};

\draw[flow]
    (2.4,12.58) -- (3.20,12.58);

\node[anchor=west, font=\sffamily\scriptsize] at (3.35,12.58)
    {operational flow};

\draw[evidence]
    (5.70,12.58) -- (6.50,12.58);

\node[anchor=west, font=\sffamily\scriptsize] at (6.65,12.58)
    {certificate / evidence};

\draw[governance]
    (9.45,12.58) -- (10.25,12.58);

\node[anchor=west, font=\sffamily\scriptsize] at (10.40,12.58)
    {governance / supervision};

\node[localstep] at (14.65,12.58)
    {};

\node[anchor=west, font=\sffamily\scriptsize] at (14.95,12.58)
    {local operation};

\draw[reference]
    (18.0,12.58) -- (18.8,12.58);

\node[anchor=west, font=\sffamily\scriptsize] at (18.95,12.58)
    {rules / reference data};

% ------------------------------------------------------------------
% Layer boxes
% ------------------------------------------------------------------

\draw[rounded corners=4pt, thick]
    (0.25,8.90) rectangle (17.45,12.2);

\node[
    anchor=west,
    font=\sffamily\bfseries,
    fill=white,
    inner sep=2pt
] at (0.95,11.9)
    {Compliance and supervisory layer};

\draw[rounded corners=4pt, thick]
    (18.55,8.90) rectangle (23.95,12.2);

\node[
    anchor=west,
    font=\sffamily\bfseries,
    fill=white,
    inner sep=2pt,
    align=left
] at (18.8,11.7)
    {Optional issuer-maintained\\audit record component};

\draw[rounded corners=4pt, thick]
    (0.25,5.60) rectangle (23.95,8.40);

\node[
    anchor=west,
    font=\sffamily\bfseries,
    fill=white,
    inner sep=2pt
] at (0.95,8.13)
    {Non-custodial asset transfer layer};

\draw[rounded corners=4pt, thick]
    (0.25,2.3) rectangle (23.95,5.1);

% ------------------------------------------------------------------
% Compliance and supervisory layer
% ------------------------------------------------------------------

\node[actor] (registries) at (7,9.7)
    {Regulatory Registries};

\node[actor] (authority) at (11.7,11.45)
    {Regulatory Authority};

\node[actor] (cas) at (11.7,9.7)
    {Compliance Attestation Service (CAS)};

% ------------------------------------------------------------------
% Optional issuer-maintained audit record component
% ------------------------------------------------------------------

\node[store] (audit) at (21.20,10.1)
    {Audit Records};

% ------------------------------------------------------------------
% Non-custodial asset transfer layer
% ------------------------------------------------------------------

\node[actor] (alice) at (2.3,7.0)
    {Alice's Wallet};

\node[actor] (bob) at (11.7,7.0)
    {Bob's Wallet};

\node[actor] (processor) at (16.4,7.0)
    {Integrity Provider};

\node[store] (ledger) at (21.20,7.0)
    {Oblivious Ledger};

% ------------------------------------------------------------------
% Regulated financial institutions
% ------------------------------------------------------------------

\node[actor] (fiAcq) at (2.3,3.2)
    {Alice's Bank\\(withdrawal)};

\node[actor] (issuer) at (7,3.2)
    {Issuer};

\node[actor] (fiRed) at (11.7,3.2)
    {Bob's Bank\\(redemption)};

% ------------------------------------------------------------------
% Standing governance, evidence, and reference-data relations
% ------------------------------------------------------------------

% Relation A:
% The regulatory authority publishes and maintains registry versions
% and policy settings.
\draw[governance]
    (authority)
    -|
    node[left] {(A)}
    (registries);

% Relation B:
% The regulatory authority authorizes or supervises CAS.
\draw[governance]
    (authority.south)
    --
    node[left] {(B)}
    (cas.north);

% Relation C:
% The regulatory authority may supervise or audit issuer-maintained
% audit records.
\draw[governance]
    (authority)
    --
    (16.4,11.45)
    |-
    node[left] {(C)}
    (audit);

% Relation E:
% Regulatory registries provide the authoritative rules and reference
% datasets used by CAS when performing compliance evaluation.
\draw[reference]
    (registries.east)
    --
    node[above] {(E)}
    (cas.west);

% ------------------------------------------------------------------
% Relation D
% ------------------------------------------------------------------

\draw[thick,dashed,->]
    (fiRed)
    --
    node[above] {(D)}
    (issuer);

\draw[evidence]
    (issuer)
    -- (7,4.8)
    -- (23.45,4.8)
    -- (23.45,10.1)
    -- (audit);

% ------------------------------------------------------------------
% Main operational sequence
% ------------------------------------------------------------------

% Step 1:
% Alice's wallet sends the token-withdrawal request to her financial
% institution.
\draw[flow]
    (alice.250)
    --
    node[left,yshift=-2.5em] {(1)}
    (fiAcq.110);

% Alice's financial institution authorizes the request and instructs
% the issuer. This is the second communication represented by step 1.
\draw[flow]
    (fiAcq.350)
    --
    node[below] {(1)}
    (issuer.190);

% Step 2:
% The issuer delivers A_0 to Alice's wallet.
\draw[flow]
    (issuer.170)
    --
    node[above] {(2)}
    (fiAcq.10);

\draw[flow]
    (fiAcq.70)
    --
    node[right,yshift=-2.5em] {(2)}
    (alice.290);

\node[
    anchor=west,
    font=\sffamily\bfseries,
    fill=white,
    inner sep=2pt
] at (0.95,4.83)
    {Regulated financial institutions};

% Steps 3, 6, and 12 are local operations performed inside Bob's Wallet.
% Rounded-square tags are attached directly to the wallet boundary so that
% they are visually distinct from numbered inter-component arrows.

\node[localstep] at ([xshift=0.38cm]bob.north west)
    {3};

\node[localstep] at ([xshift=0.38cm]bob.south west)
    {6};

\node[localstep] at ([xshift=-0.38cm]bob.south east)
    {12};

% Step 4:
% Bob submits F_j to CAS.
\draw[flow]
    (bob.110)
    --
    node[left,yshift=-1em] {(4)}
    (cas.250);

% Step 5:
% CAS returns alpha_j to Bob after evaluating F_j against the
% authoritative regulatory rules and reference datasets.
\draw[evidence]
    (cas.290)
    --
    node[right,yshift=-1em] {(5)}
    (bob.70);

% Step 7:
% Bob sends h(F_j) to Alice.
\draw[flow]
    (bob.170)
    --
    node[above] {(7)}
    (alice.10);

% Step 8:
% Alice returns the resulting update information to Bob.
\draw[flow]
    (alice.350)
    --
    node[below] {(8)}
    (bob.190);

% Step 9:
% Bob submits the update material to the integrity provider.
\draw[flow]
    (bob.350)
    --
    node[below] {(9)}
    (processor.190);

% Step 10:
% The integrity provider validates and commits through the finality layer.
\draw[flow,<->]
    (processor.east)
    --
    node[above] {(10)}
    (ledger.west);

% Step 11:
% The integrity provider returns the commitment proof to Bob.
\draw[evidence]
    (processor.170)
    --
    node[above] {(11)}
    (bob.10);

% Step 13:
% Bob presents A_j to his financial institution for redemption.
\draw[flow]
    (bob)
    --
    node[left,yshift=-2.5em] {(13)}
    (fiRed);

% ------------------------------------------------------------------
% Event list box
% ------------------------------------------------------------------

\draw[rounded corners=0pt, thin]
    (0.25,-2.0) rectangle (23.95,2.0);

\node[anchor=west, font=\sffamily\bfseries]
    at (0.95,1.7)
    {Main sequence};

\node[anchor=west, font=\sffamily\bfseries]
    at (15.9,1.7)
    {Standing relations};

\node[eventtext, text width=7.2cm] at (0.95,1.3) {
1. Alice's wallet initiates a token-withdrawal request, which her financial institution authorizes.\\
2. Issuer delivers $A_0$ to Alice, who holds it in self-custody.\\
3. When Bob decides to request a payment from Alice, his wallet constructs update vector $F_j$.\\
4. Bob submits $F_j$ to CAS.\\
5. After successful evaluation, CAS returns $\alpha_j$ only.\\
6. Bob embeds compliance signature $\alpha_j$ in message field $u_j$ of $F_j$.\\
7. Bob sends $h(F_j)$ to Alice.
};

\node[eventtext, text width=7.2cm] at (8.30,1.3) {
8. Alice uses $h(F_j)$ to create update $U_j$ using the signature key corresponding to $k_j$ and sends this along with her context $u^s_j$ and the previous state of the asset $A_{j-1}$ to Bob.\\
9. Bob submits a hash of the update and its signature along with $k_j$ to the integrity provider.\\
10. Integrity provider validates and commits.\\
11. Commitment proof returns.\\
12. Bob keeps asset $A_j$ in self-custody.\\
13. Bob may present $A_j$ to his financial institution for redemption.
};

\node[eventtext, text width=7.7cm] at (15.9,1.3) {
A. Authority publishes registry versions and policy settings.\\
B. Authority authorizes or supervises CAS.\\
C. Auditors or authorities may audit issuer-maintained records.\\
D. Redemption institutions and exchange services may submit compliance certificates to the Issuer, which records the relevant evidence in the issuer-maintained audit record.\\
E. Regulatory Registries constitute the authoritative source of rules and reference datasets used by CAS to evaluate proposed asset updates.
};

\end{tikzpicture}

\caption{Schematic event sequence for the proposed reference architecture,
showing asset acquisition, self-custodial holding, compliance attestation,
update validation, continued self-custodial holding, possible redemption,
and optional issuer-maintained audit records.}
\label{fig:general-reference-architecture}

\end{figure}
\end{landscape}

The following subsections disaggregate this integrated view. Section~\ref{sec:actors-and-roles} identifies the actors and roles responsible for the principal functions shown in the architecture, while Section~\ref{sec:components} describes the components through which asset state, validation, compliance evaluation, and auditability are implemented.

\subsection{Actors \& Roles}
\label{sec:actors-and-roles}

We identify a set of actors and roles that perform distinct functions across asset issuance, validation, compliance enforcement, and value transfer within the architecture.

\subsubsection{Issuer (Minter)}

The issuer is responsible for asset creation, refreshment, and destruction. It generates assets with verifiable initial provenance, authorizes reissuance where required by protocol rules, and decommissions assets upon redemption or invalidation under defined conditions. 

Refreshment (or reissuance) refers to a privacy-preserving process in which an existing asset held in self-custody is extinguished and replaced with a newly issued asset of equivalent value upon presentation of proofs attesting to its validity and policy compliance \cite{friolo_goodell_toliver_nakib_private_payments_zk_reissuance_2024}.  Because the replacement asset corresponds to money already in circulation, the process does not increase the monetary base. Instead, it exchanges an asset representing money that has already been spent for money that can be spent in the future, while preventing the new instance from carrying forward a linkable record of the previous asset's transaction history.

\subsubsection{Payment Processor}

Following Goodell, Toliver, and Nakib \cite{goodell_toliver_nakib_scalable_payments_2021}, the payment processor in our architecture adopts the relay-based design proposed in their scalable payment framework. Its function is limited to receiving and aggregating cryptographic commitments derived from hashes of asset updates and recording evidence of those commitments within the supporting infrastructure. Because it is oblivious and therefore does not receive or inspect the contents of the underlying updates, it makes no determination regarding their structural, cryptographic, economic, or regulatory validity.

The payment processor therefore performs a narrowly defined notarial and data-aggregation function rather than acting as a conventional financial intermediary. It does not undertake customer identification, transaction monitoring, sanctions screening, or compliance verification. These functions remain at the edges of the transaction, where they are performed by the relevant regulatory authorities, designated institutions, or transaction counterparties.

Assets may nevertheless be transferred and their commitments anchored without carrying sufficient compliance evidence. Transaction counterparties have an incentive to require such evidence because the asset's future acceptability and redeemability may depend on it. This functional separation prevents regulatory requirements designed for financial intermediaries from being unnecessarily extended to actors performing a materially different communications and data-aggregation function.

\subsubsection{Regulatory Authority (Auditor)}

The regulatory authority establishes applicable compliance requirements and exercises supervisory oversight. It oversees both traditional compliance infrastructure and the digital compliance infrastructure introduced by our proposed architecture. The latter consists of three components:
\begin{itemize}
    \item Compliance attestation services (see Section~\ref{sec:compliance-attestation-services})
    \item Regulatory registries (see Section~\ref{sec:regulatory-registries})
    \item Audit logs (see Section~\ref{sec:audit-records})
\end{itemize} The authority conducts audits and ensures that compliance conditions are correctly implemented and independently verifiable across both environments.

\subsubsection{Senders and Receivers}

Senders and receivers are users who hold and transfer assets. Assets may be held in self-custody or through custodial arrangements, depending on user preference and regulatory configuration. Senders construct proposed state updates and obtain required compliance certificates prior to submission. Receivers accept assets whose state transitions have been validated and recorded.

\subsubsection{Financial Institutions}

Licensed financial institutions facilitate interaction between the self-custodial digital asset environment and established monetary systems. They debit conventional monetary value in exchange for issued tokens and process redemption of these tokens back into traditional monetary forms, such as cash and bank deposits, subject to verification that regulatory conditions are satisfied.

\subsubsection{Exchange Services}

Cross-border or multi-currency transfers may require exchange services. A self-custodied token denominated in one currency cannot ordinarily be redeemed by a financial institution that operates only in another currency zone. Exchange services can intermediate between currency domains, match countervailing flows, or provide liquidity. Their participation introduces additional compliance predicates relating to jurisdiction, exchange rate, licensing, reporting, and redemption endpoints.

\subsection{Components}
\label{sec:components}

We identify the following core components to formalize the functional structure of the architecture and clarify how asset state, validation, compliance, and auditability are implemented within the protocol.

\subsubsection{USO Asset}

The asset component of the architecture is based on the USO model developed by Goodell et al.~\cite{goodell_toliver_nakib_scalable_payments_2021}. In this model, a digital asset is not represented as an account balance maintained on a ledger. Instead, the asset is a portable digital object that may be held directly by the user and transferred through signed updates to its state. The asset carries the information required to verify its valid progression, while the ledger does not manage the asset itself or process transaction contents. The ledger's role is limited to providing evidence that commitments to asset updates have been incorporated into an authoritative history.

The relevance of the USO model for this paper lies in the structure of the asset itself and in the form that asset updates take.  Adapting the notation of Lynham et al.~\cite{lynham_sark_2025}, an asset update proposed by a prospective new controller (i.e., a recipient) at sequence number $j$ is represented as
\begin{equation}
F_j \leftarrow u_j \,||\, h(G_{j,0}, k_{j+1}),
\end{equation}
where $u_j$ is an open message field, $G_{j,0}$ is a reference to a root of an oblivious ledger $L$, and $k_{j+1}$ is the next controller's one-time verification key. To effect a transfer, the current controller (i.e., the sender) authorizes the update by signing the hash of the update vector, where $u^s_j$ is a field representing transaction context provided by the current controller and $A_{j-1}$ is the previous state of the asset:
\begin{equation}
\label{e:U_j}
U_j \leftarrow u^s_j\,||\,h(F_j)\,||\,h(A_{j-1}), s(h(u^s_j\,||\,h(F_j)\,||\,h(A_{j-1})),
k_j).
\end{equation}
The updated asset is then obtained by appending the signed update, new anchor, and new key to the prior asset state:
\begin{equation}
A_j \leftarrow A_{j-1} \,||\, (U_j,G_{j,0},k_{j+1}). \label{e:A_j}
\end{equation}

For this paper, the most relevant feature of the USO model is that each asset update contains an open message field, denoted $u_j$. This message field is not constrained to any particular use or purpose in the baseline model and may be left blank. The proposed architecture uses that field as the location where embedded compliance is introduced into the asset state. This allows compliance-relevant evidence generated before or during each transfer to travel with the asset across its lifecycle, so that an auditor can later examine the asset's state history and verify whether prior transfers satisfied the applicable compliance conditions, without requiring the ledger itself to record or inspect transaction contents.

In the proposed architecture, $u_j$ is populated with a structured compliance envelope. In a retail payment, the envelope may contain, for example, a \texttt{purchase-context} message recording basic information about the transaction, such as when and where it occurred and what type of merchant was involved. It may also include a \texttt{merchant-credential} message, linking the update to evidence that the merchant is authorised or recognised for that transaction type, and a \texttt{redeemer-authorisation} message identifying a specific financial institution authorised to redeem or receive the asset, for example through a Legal Entity Identifier (LEI). Other messages may point to tax-relevant records or audit entries, meaning records in an append-only audit log that allow a supervisory authority to verify that a compliance-relevant event was recorded. The envelope may also contain the compliance signature issued by an authorised Compliance Attestation Service (see Section~\ref{sec:compliance-attestation-services}), attesting that the asset update was evaluated and rendered compliant with the applicable rule set.

Because $u_j$ is part of the update vector, the compliance envelope becomes part of the signed asset update and therefore part of the asset's evolving state. An auditor can later examine the asset's state history, verify the signatures and attestations associated with each relevant update, check the applicable registry versions and policy references, and determine whether the asset evolved through state transitions that satisfied the required compliance conditions. Compliance can therefore be verified deterministically from the asset's own provenance and embedded attestations, without requiring the ledger to record or inspect transaction contents.

As shown in Figure~\ref{fig:compliance-envelope-rust}, the USO message field may be represented as a vector of update messages. One element of that vector may be a compliance envelope, which contains the compliance-relevant messages and attestations associated with the update.

\begin{center}
\begin{minipage}[t]{0.48\linewidth}
\begin{lstlisting}[style=compactrust]
pub enum UpdateMessage {
    ComplianceEnvelope(
        ComplianceEnvelope
    )
}

pub struct ComplianceEnvelope {
    pub messages: Vec<ComplianceMessage>,
}

pub enum ComplianceMessage {
    PurchaseContext {
        purchase_timestamp: String,
        jurisdiction: String,
        merchant_category_code: String,
        receipt_reference: String,
    },
    MerchantCredential {
        merchant_id: String,
        credential_issuer: String,
        merchant_credential_commitment: String,
    },
    RedeemerAuthorization {
        institution_lei: String,
        lei_scheme: String,
        redemption_jurisdiction: String,
        authorization_commitment: String,
    },
    TaxContext {
        tax_jurisdiction: String,
        tax_registration_commitment: String,
        fiscal_receipt_hash: String,
        tax_category: String,
    },
\end{lstlisting}
\end{minipage}
\hfill
\begin{minipage}[t]{0.48\linewidth}
\begin{lstlisting}[style=compactrust]

    ComplianceAttestation {
        attestation_issuer: String,
        result: String,
        signed_digest: String,
        signature: String,
    },
    AuditReference {
        audit_log_id: String,
        audit_entry_hash: String,
        supervisory_authority: String,
        recorded_at: String,
    },
}

pub struct AssetUpdateVector {

    pub messages: Vec<UpdateMessage>,

    pub ledger_root: String,

    pub next_controller_key: String,
}


let messages = vec![
    UpdateMessage::ComplianceEnvelope(
        compliance_envelope
    ),
];
\end{lstlisting}
\end{minipage}

\captionof{figure}{Simplified Rust representation of a USO asset update whose open message field contains a vector of update messages, one of which is a compliance envelope.}
\label{fig:compliance-envelope-rust}
\end{center}

\subsubsection{Compliance Attestation Services}
\label{sec:compliance-attestation-services}

The Compliance Attestation Service (CAS) is a service that evaluates whether a proposed update vector $F_j$ satisfies the applicable compliance predicate set $C_j$ under a specified registry version $v$, as defined in Section~\ref{sec:regulatory-registries}. Upon successful evaluation, CAS produces a cryptographically verifiable attestation $\alpha_j$ bound to the update context.

In the notation of Lynham et al.~\cite{lynham_sark_2025}, $F_j$ denotes the update vector, $U_j$ denotes the signed update, and $A_j$ denotes the updated asset state. CAS evaluates the proposed update vector before the controller produces the signed update $U_j$. However, because the compliance attestation is itself carried inside the compliance envelope embedded in $u_j$, CAS cannot sign a representation of $F_j$ that already contains $\alpha_j$. Instead, CAS signs a canonical pre-attestation representation of the proposed update vector. Let $F_j^{-}$ denote this representation, obtained from $F_j$ by leaving the compliance-attestation field empty or excluding it from the signed digest.

Let $C_j = \{r_1, \ldots, r_m\}$ denote the set of compliance constraints applicable to $F_j^{-}$, where each $r_i$ references a regulatory rule identifier contained in registry $R_v$.

If all predicates in $C_j$ are satisfied for $F_j^{-}$, CAS generates an attestation
\begin{equation}
\alpha_j =
\mathsf{Sign}_{k_{\mathsf{CAS}}}
(H(F_j^{-} \parallel v \parallel C_j \parallel \mathsf{result}_j)), \label{e:alpha_j}
\end{equation}
where $H(\cdot)$ is a cryptographic hash function, $k_{\mathsf{CAS}}$ is the signing key of the Compliance Attestation Service, and $\mathsf{result}_j$ records the outcome of the compliance evaluation.

The attestation $\alpha_j$ is then inserted into the compliance envelope carried by $u_j$. The controller may then sign the resulting update vector to produce the signed update $U_j$, which is appended to the prior asset state to form $A_j$. Verification does not re-execute the compliance checks. It confirms that CAS signed the canonical pre-attestation representation $F_j^{-}$ under registry version $v$, that the applicable predicate set $C_j$ was used, and that the resulting attestation corresponds to the compliance envelope carried in the signed update $U_j$.

\paragraph{Information visible to the CAS.} The CAS is contacted by the recipient before
the payer takes part in the transaction, so it learns nothing about the payer. The request
carries recipient-side context only: the recipient's authorisation credential, the amount
and category of the transfer, and the recipient's commitment to the proposed update.
Section~\ref{sec:uc-1} gives an example of this in a retail payment.

A related design question concerns the scope of the compliance attestation itself. A compliance signature may attest only to transaction-level facts, or it may also be linked to structural elements of the asset update in which those facts appear. The latter approach addresses the risk that a valid compliance signature could be reused across another update carrying similar transaction information. The following subsection describes this update-specific binding option and explains how it can strengthen the relationship between compliance evidence and the asset transition to which it refers.

\subsubsection{Optional Update-Specific Binding of Compliance Attestations}
\label{sec:compliance-attestation-binding}

The scope of the compliance attestation can be defined with different levels
of specificity. A minimal design may treat the attestation as a signature over
transaction-level facts, such as the recipient's authorization, the permitted
amount, applicable transaction limits, or tax-related information. This provides
evidence that those facts were evaluated, but it does not by itself determine
how tightly the resulting signature is linked to the particular asset update in
which it later appears.

A stricter design can bind the attestation to structural elements of the update
context. In addition to the transaction facts contained in the compliance
envelope, the signed material can include the ledger reference \(G_{j,0}\) and
the recipient's one-time verification key \(k_{j+1}\). The ledger reference identifies
the integrity provider through which the next update is expected to be
registered, while \(k_{j+1}\) identifies the recipient-controlled key under
which the next asset state will be held. In this sense, \(k_{j+1}\) functions
as the cryptographic address of the next update, rather than as a transaction
number or a fixed position within a block.

This design choice addresses a replay or substitution risk. If a compliance
signature attests only to general transaction facts, the same signature could
potentially be copied into another update containing similar information,
falsely suggesting that the second update had also received compliance
approval. By also covering \(G_{j,0}\) and \(k_{j+1}\), the signature links the
attested facts to a particular integrity layer and to the specific
recipient-controlled position at which the next asset update is expected to
occur.

The resulting attestation therefore applies to a defined update context rather
than operating as a transferable compliance approval across unrelated
transactions. This does not change the role of the Compliance Attestation
Service or the integrity provider. CAS still evaluates the proposed update and
issues the attestation, while the integrity provider remains concerned with
structural validity and commitment. The additional effect is that the
compliance evidence carried in \(u_j\) becomes more tightly associated with the
asset transition for which it was produced.

Update-specific binding should therefore be understood as a complementary
design option within the broader architecture. It is especially useful where
the policy objective is to prevent replay, substitution, or reuse of valid
compliance signatures across otherwise similar asset transitions, without
requiring the ledger itself to inspect transaction contents.
\subsubsection{Compliance Certificates (or Compliance Signatures)}
\label{sec:compliance-certificates}

A compliance certificate is a cryptographically signed message issued by a recognized Regulatory Authority or its delegate. It states that a specified asset, transfer, update, or provenance chain satisfies a defined compliance condition. Within the proposed architecture, such a certificate is treated as verifiable evidence that the certified condition has been met. The certificate does not itself perform compliance enforcement; rather, it provides the cryptographic proof on which protocol components, wallets, custodial institutions, or redemption interfaces may rely when deciding whether to accept, process, or convert an asset.

We use the term \emph{compliance certificate} generically to refer to signed compliance evidence produced by an authorized compliance service. This includes both transfer-level attestations and broader certificates relating to the compliance status of an asset state or its provenance chain. Two forms are relevant for the present architecture.

\begin{itemize}

\item[(i)] A transfer-level attestation $\alpha_j$, certifying that a proposed update vector at hop $j$ satisfies the applicable compliance predicates under a specified registry version $v$. Formally, $\alpha_j$ is the signature produced by CAS over a hash of the canonical pre-attestation representation $F_j^{-}$, the registry version $v$, the applicable predicate set $C_j$ and the outcome of the evaluation, as defined in Eq.~\ref{e:alpha_j} (Section~\ref{sec:compliance-attestation-services}). Verification of $\alpha_j$ consists of validating the signature under the corresponding verification key, confirming that the signed hash corresponds to the proposed update attempt, and checking consistency between $v$ and the applicable enforcement configuration. Once embedded in the compliance envelope associated with the update, $\alpha_j$ provides cryptographic evidence that the transfer satisfied the applicable compliance predicates at hop $j$.

\item[(ii)] A convertibility certificate $\pi(A_j)$, certifying that asset state $A_j$ and its provenance chain satisfy the applicable compliance predicates for conversion into another monetary form. This certificate serves as evidence that the asset is eligible for conversion into account-based funds, cash, or other conventional monetary forms, and is verified by the redeeming institution prior to executing conversion.

\end{itemize}

Two kinds of evidence can enter the compliance envelope of an update: real-time attestations, which the CAS signs when the transfer is attempted, as in item~(i); and certificates that already exist, such as a merchant's authorisation credential, which are added to the envelope at transfer time and checked by the CAS rather than re-issued. The architecture allows either or both, and the enforcement policy of Section~\ref{sec:core-protocol-rules} states which are required for a given transfer.

\subsubsection{Regulatory Registries}
\label{sec:regulatory-registries}

Regulatory registries are structured, authoritative data repositories containing
machine-readable policy parameters and reference datasets used in compliance
evaluation. Registries publish and version compliance predicates together with
their associated reference data, enabling dynamic rule maintenance without
modification of the core protocol.

Formally, let $R$ denote a regulatory registry. We model $R$ as a versioned
structure
\begin{equation}
R = (C, D, v),
\end{equation}
where:

\begin{itemize}
    \item $C = \{ c_1, c_2, \dots, c_n \}$ is a finite set of compliance predicates;
    \item $D = \{ D_1, D_2, \dots, D_m \}$ is a collection of reference datasets used in evaluating those predicates;
    \item $v$ denotes the registry version identifier.
\end{itemize}

Each predicate $c_i \in C$ is evaluated over the proposed update vector $F_j$,
possibly with reference to one or more datasets in $D$. At this stage, $F_j$
refers to the update vector submitted for compliance evaluation, not to the
signed update $U_j$ or to the updated asset state $A_j$. Formally,
\begin{equation}
c_i : F_j \times D \rightarrow \{ \text{true}, \text{false} \}.
\end{equation}

A simple example illustrates the relation between predicates and datasets. Let
$c_{\mathsf{merchant}}$ denote a merchant-admissibility predicate. This predicate
may evaluate to true only if the merchant credential contained in the compliance
envelope includes a valid Legal Entity Identifier (LEI), the LEI appears in an
authorised-merchant dataset, and the same LEI does not appear in a relevant
blacklist, sanctions list, or suspension dataset. Similarly, a tax-context
predicate may evaluate to true only if the purchase context includes a fiscal
receipt reference and the merchant category is mapped to an accepted tax category
in the applicable tax dataset. These examples show that compliance predicates
depend on versioned reference datasets: the rule defines the condition, while the
dataset supplies the authoritative facts against which the condition is evaluated.

The registry thus serves as the authoritative source of both normative conditions
(predicates) and the operational data required for their evaluation. Updates to
compliance requirements are effected by publishing a new registry version $v'$
without modification to the underlying asset protocol.

\subsubsection{Audit Records}
\label{sec:audit-records}

In conventional electronic payment systems, compliance enforcement is supported
by custodial visibility over payment flows. Card payments, interbank transfers,
and e-money transfers are processed by institutions that maintain balances,
execute movements between accounts, and record account-level activity. Competent
authorities may then require these regulated intermediaries to disclose
information about accounts, balances, counterparties, and payment activity.

This model provides a basis for supervisory access, but not for structural
payment privacy. Because enforcement depends upon the observation or
reconstruction of user payment flows, the privacy of users remains contingent
upon institutional access controls rather than protected by the architecture
itself. A privacy-preserving compliance architecture should therefore avoid
making general payment-flow visibility the basis of enforcement.

The architecture proposed here explores a different design direction. Rather
than treating payment intermediation as the source of compliance visibility,
compliance evidence is generated in relation to asset-state transitions and may
be embedded within, or associated with, the evolving asset state. The relevant
design question is how such evidence can be preserved for later verification
without requiring a comprehensive record of user payment flows.

One possible way to address this question is through an optional
issuer-maintained audit record. Under this design, the issuer, or an
issuer-operated service, maintains an append-only record of selected
compliance-relevant asset lifecycle events. The record is not part of the
oblivious ledger or integrity layer, and it is not a substitute for the
Compliance Attestation Service. Instead, it functions as a complementary
evidentiary component: a place where compliance certificates, or references to
such certificates, may be recorded for later inspection by an authorised
auditor.

The rationale for associating this component with the issuer is that the issuer
is already responsible for asset creation, refreshment, and destruction. It is
therefore a natural institutional point at which lifecycle evidence relating to
\texttt{issuance}, \texttt{redemption}, \texttt{burn}, \texttt{exchange}, or
\texttt{refresh} functions may be collected without
requiring the integrity layer to observe transaction contents. The record would
track asset-level compliance evidence, not people. It need not contain a
complete history of user identities, counterparties, or payment flows.

Compliance evidence may be contributed to this record at different points in the
asset lifecycle. During transfer, a Compliance Attestation Service may issue a
transfer-level attestation \(\alpha_j\), which is embedded in the compliance
envelope carried by \(u_j\).  At later lifecycle events, such as
\texttt{redemption}, \texttt{burn}, \texttt{exchange}, or \texttt{refresh},
institutions providing redemption and exchange services may submit additional
compliance certificates to the issuer-maintained audit record, creating
auditable evidence that the relevant asset transition, or the provenance chain
on which it depends, satisfied the applicable compliance predicates.

Formally, an audit record may be represented as an append-only structure

\begin{equation}
\mathcal{L} = \{\ell_1, \ell_2, \ldots, \ell_n\},
\end{equation}
where each entry is given by
\begin{equation}
\ell_k = (e_k, t_k, h_k, \gamma_k).
\end{equation}
Here, \(e_k\) denotes a compliance-relevant lifecycle event, \(t_k\) denotes a
timestamp, \(h_k\) denotes a cryptographic commitment binding the entry to the
relevant asset context, and \(\gamma_k\) denotes a compliance certificate,
certificate reference, or commitment to such a certificate. The record would be
append-only and integrity-protected, so that entries could not be modified
without detection.

A useful analogue is provided by burn-and-reissuance mechanisms in
privacy-preserving payment systems. Friolo et al. describe a protocol in which
zero-knowledge proofs are combined with an audit log so that a payer can prove
that newly issued assets were created through an ``atomic process wherein assets
of equal value were destroyed,'' without revealing the specific circumstances of
creation \cite{friolo_goodell_toliver_nakib_private_payments_zk_reissuance_2024}.
They also characterise the relevant recirculation function as one in which
tokens are ``burned'' and ``minted'' in equal measure
\cite{friolo_goodell_toliver_nakib_private_payments_zk_reissuance_2024}.

This analogy illustrates the possible role of audit records in the present
architecture. A \texttt{burn} event refers to the retirement, destruction, or
decommissioning of a prior asset state, while \texttt{refresh} or
\texttt{reissuance} refers to the creation of a new asset state that replaces
it. An issuer-maintained audit record could preserve evidence that the new state
is supported by a valid prior event, without exposing the user payment flows
through which the asset circulated.

This mechanism remains optional at the architectural level. Under a stricter
policy configuration, \texttt{redemption}, \texttt{refresh}, or
\texttt{reissuance} may require a compliance certificate linked to a recorded
burn, redemption, or exchange event.  Under a less strict configuration, the
asset may be redeemed, decommissioned, or reissued without additional
audit-record certification, provided that the applicable policy permits it.

The issuer-maintained audit record should therefore be understood as a
complementary design option. It supports independent supervisory review and
evidentiary traceability by preserving selected compliance-relevant asset
lifecycle evidence, while leaving the oblivious ledger focused on integrity and
avoiding the creation of a comprehensive ledger of user payment histories.

\section{Core Protocol Rules}
\label{sec:core-protocol-rules}

\subsection{What Constitutes a Protocol}

A protocol specifies permitted actions and the conditions under which those actions are valid. It defines correctness independently of any particular system realization. Multiple architectures may implement the same protocol, provided they satisfy its constraints.
Certain definitions and constraints may be fixed at the protocol level. These constraints delimit the class of admissible systems and determine the scope within which architectural variation is permitted.

\subsection{Compliance as a Protocol-Level Constraint}

In conventional payment systems, auditability depends on records maintained by intermediaries acting as custodians, introducing reliance on opaque institutional processes. As a result, such systems have a limited ability to provide uniform guarantees of consistent and verifiable compliance enforcement across participants and transactions.

The approach taken here repositions compliance as part of the protocol’s internal logic: a proposed asset update (e.g. a token transfer) is considered valid only if it satisfies the applicable compliance conditions. This design allows enforcement to occur deterministically at the level of the protocol.

This section formalizes protocol-level constraints governing asset state transitions, ensuring that such transitions occur in accordance with rules defined by the relevant regulatory authority.

We propose a core protocol rule: asset updates are valid only if the applicable compliance constraints are satisfied and a corresponding compliance certificate is embedded in the successor state.

Formally, let $A_{j-1}$ denote the asset state immediately prior to update $j$, and let $F_j$ denote the proposed update vector. Let
\begin{equation}
C_j = \{r_1,\dots,r_m\}
\end{equation}
denote the set of compliance constraints applicable to the update, where each $r_i$ references a rule identifier in an authoritative regulatory registry under the relevant jurisdiction.

Define the protocol-level predicate $C(A_{j-1}, F_j) \in \{0,1\}$ by
\begin{equation}
C(A_{j-1}, F_j) = 1 \iff \forall r_i \in C_j,\; r_i \text{ is satisfied.}
\end{equation}

A compliance certificate attesting to the admissibility of the proposed state transition is issued if and only if
\begin{equation}
C(A_{j-1}, F_j) = 1.
\end{equation}
If
\begin{equation}
C(A_{j-1}, F_j) = 0,
\end{equation}
then no compliance certificate is issued, and no successor state $A_j$ can be rendered compliant under the protocol.

Compliance is therefore enforced by protocol-level admissibility conditions governing state evolution, rather than by institutional review after transaction  execution.

\subsection{Enforcement Points}

In our model, compliance enforcement operates at three points in the asset lifecycle:
\begin{itemize}
    \item \textbf{Withdrawal:} prior to asset issuance, performed by the financial institution that would facilitate the conversion.
    \item \textbf{Transfer:} asset state transitions between controllers.
    \item \textbf{Redemption:} token conversion back into legacy monetary form.
\end{itemize}

These correspond to two rule classes:
\begin{itemize}
    \item \textbf{Transfer-Level Rules:} governing ownership updates through state transitions.
    \item \textbf{Convertibility Rules:} governing the admissibility conditions under which self-custodial asset states may be converted into institutionally mediated monetary forms.
\end{itemize}

For an examination of how these protocol-level rules are instantiated within a retail payment system implementing our reference architecture, see Section~\ref{sec:use-cases}.

\subsection{Transfer-Level Rules}

Transfer-level rules govern admissibility of asset state transitions. If the conditions required for a valid update are not satisfied, the authorization necessary to produce a valid successor state is not generated, and it is not possible to create an update that is considered valid.  Recipients of invalid assets will be able to recognize the asset as invalid and reject it.  Admissibility is therefore enforced by non-recognition of invalid state transitions at the point of an attempted ownership update.

In addition to outright non-execution, transfer-level rules may admit a transition while embedding compliance-relevant constraints into the resulting asset state. In such cases, the transfer is provisionally effective within the non-custodial domain, but subsequent convertibility or institutional intake may be conditioned on verification of the embedded compliance attestations.

At exit from the non-custodial domain, assets presented for redemption are subject to verification of their full state history through compliance certificates generated during prior updates. Only assets whose accumulated attestations satisfy the applicable admissibility conditions may be converted into institutionally mediated monetary form. Assets that fail such checks may be placed into temporary institutional custody pending further review.

As a result, transfer-level enforcement operates through two distinct modes:
\begin{itemize}
    \item \textbf{Non-execution:} invalid updates are not admitted and no successor state is recognized as valid without institutional review and intervention.
    \item \textbf{Conditional progression:} valid updates embed compliance constraints that govern subsequent admissibility at points of convertibility.
\end{itemize}

\subsection{Convertibility Rules}

Convertibility rules govern admissibility conditions for conversion between the non-custodial asset domain and institutionally mediated monetary form.

At withdrawal, issuance of an asset into the non-custodial domain may be conditioned on satisfaction of specified admissibility criteria defined by the applicable enforcement policy.

At redemption or custodial intake, when a user presents $A_j$ for conversion or institutional admission, the receiving institution verifies the asset’s authenticity, continuity of provenance, and the presence and validity of required compliance attestations. Upon successful evaluation, the institution generates a convertibility compliance certificate $\pi(A_j)$ attesting that the asset and its cumulative update history satisfy the applicable convertibility constraints.

Acceptance into custody is conditioned on successful verification of this certificate:
\begin{equation}
\text{CustodialAcceptance}(A_j) \iff \text{verify}(\pi(A_j))=\text{true}.
\end{equation}

Under the audit-log operational model, the certificate $\pi(A_j)$ may be committed to an append-only log accessible to the relevant supervisory authority. Depending on the adopted publication configuration, this log may be regulator-restricted or publicly verifiable.

\subsection{Security Properties}
\label{sec:security-properties}

We give informal arguments for the following properties, under a threat model in which the payment processor and integrity provider are honest-but-oblivious (they correctly commit updates but do not inspect their contents) and the Compliance Attestation Service's signing key $k_{\mathsf{CAS}}$ is uncompromised.

\textbf{Attestation unforgeability.} Since $\alpha_j$ is a signature under $k_{\mathsf{CAS}}$ over the canonical pre-attestation representation $F_j^{-}$ (Eq.~\ref{e:alpha_j}), no party other than CAS can produce a valid attestation for a non-conforming update, assuming existential unforgeability of the underlying signature scheme. A forged or copied attestation is detectable at verification, since it will not validate against $F_j^{-}$ under the claimed registry version $v$.

\textbf{Provenance integrity.} Because each signed update $U_j$ binds the hash of the prior state $h(A_{j-1})$ (Eq.~\ref{e:U_j}), and the updated state $A_j$ is defined by appending $U_j$ to $A_{j-1}$ (Eq.~\ref{e:A_j}), any retrospective modification of an earlier state is detectable by recomputing the hash chain. This gives the asset the same tamper-evidence guarantee as a hash-linked ledger, without requiring a ledger to store or process the asset itself.

\textbf{Replay resistance.} Under the optional update-specific binding described in Section~\ref{sec:compliance-attestation-binding}, $\alpha_j$ additionally covers the ledger reference $G_{j,0}$ and the next-controller key $k_{j+1}$. This prevents a valid attestation from being copied into a different update that shares similar transaction facts, since doing so would require a signature over different structural elements, which is infeasible without $k_{\mathsf{CAS}}$.

\textbf{Payer unlinkability.} Each transfer employs a fresh one-time verification key $k_{j+1}$, and the asset carries no persistent account identifier. An observer with access only to the asset's message history therefore cannot associate successive controllers with a stable identity, preserving the privacy objective motivating this work (Section~\ref{sec:background-and-related-work}) even as compliance evidence accumulates independently within $u_j$.

These are informal arguments, not a formal security analysis, which we leave to future work. They hold conditional on the stated trust assumptions and do not extend to compromise of a controller's own signature key, a scenario examined separately as an adversarial case in Section~\ref{sec:use-cases}.

\section{Use Cases}
\label{sec:use-cases}

The reference architecture supports multiple transfer paths involving self-custodial holders, custodial institutions, and exchange services. These paths differ according to whether value remains under direct user control, enters institutional custody, returns from custody to self-custodial circulation, or passes through an exchange service as part of a cross-border transfer.

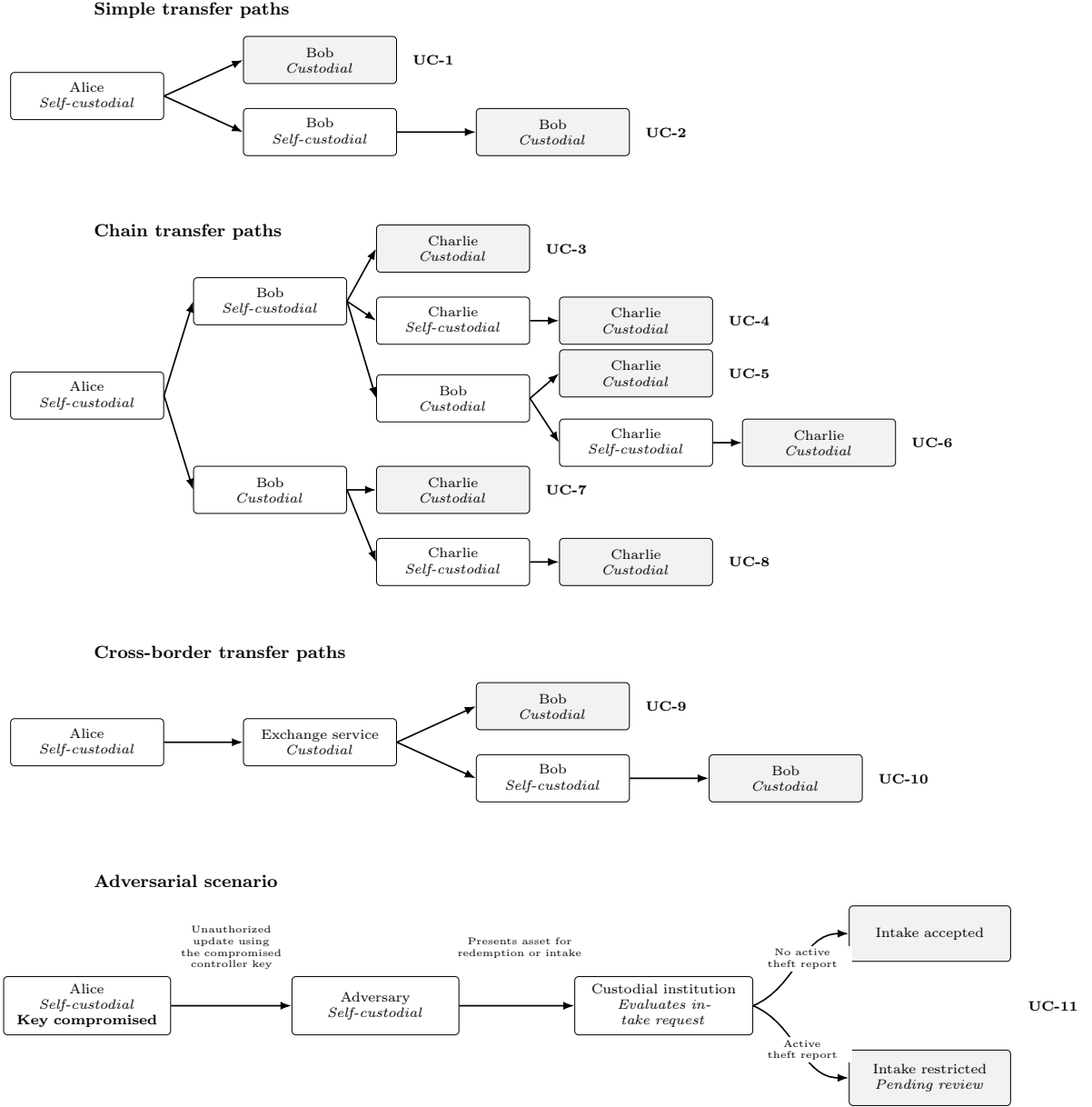
\begin{figure}[H]
\centering

% Scale the complete diagram to 96% of the text width and centre it.
% This preserves the layout while giving the figure more surrounding space.
\makebox[\textwidth][c]{
\resizebox{1.0\textwidth}{!}{
\begin{tikzpicture}[
    state/.style={
        draw,
        rounded corners=2pt,
        align=center,
        minimum height=0.85cm,
        text width=2.55cm,
        inner sep=3pt,
        font=\scriptsize
    },
    endpoint/.style={
        state,
        fill=gray!10
    },
    flow/.style={
        -{Latex[length=2mm]},
        thick
    },
    case/.style={
        font=\scriptsize\bfseries,
        anchor=west
    },
    sectiontitle/.style={
        font=\small\bfseries,
        anchor=west
    },
    edgeannotation/.style={
        font=\tiny,
        align=center,
        text width=2.8cm,
        fill=white,
        inner sep=1.2pt
    },
    branchannotation/.style={
        font=\tiny,
        align=center,
        text width=1.7cm,
        fill=white,
        inner sep=1.2pt
    }
]

% ================================================================
% SIMPLE TRANSFER PATHS
% ================================================================

\node[sectiontitle] at (0,1.55) {Simple transfer paths};

\node[state] (s-alice) at (0,0) {
    Alice\\
    \textit{Self-custodial}
};

\node[endpoint] (s-bob-c) at (4.2,0.65) {
    Bob\\
    \textit{Custodial}
};

\node[state] (s-bob-nc) at (4.2,-0.65) {
    Bob\\
    \textit{Self-custodial}
};

\node[endpoint] (s-bob-c2) at (8.4,-0.65) {
    Bob\\
    \textit{Custodial}
};

\draw[flow] (s-alice.east) -- (s-bob-c.west);
\draw[flow] (s-alice.east) -- (s-bob-nc.west);
\draw[flow] (s-bob-nc.east) -- (s-bob-c2.west);

\node[case] at (5.75,0.65) {UC-1};
\node[case] at (9.95,-0.65) {UC-2};

% ================================================================
% CHAIN TRANSFER PATHS
% ================================================================

\node[sectiontitle] at (0,-2.45) {Chain transfer paths};

\node[state] (c-alice) at (0,-5.4) {
    Alice\\
    \textit{Self-custodial}
};

\node[state] (c-bob-nc) at (3.3,-3.7) {
    Bob\\
    \textit{Self-custodial}
};

\node[state] (c-bob-c-direct) at (3.3,-7.1) {
    Bob\\
    \textit{Custodial}
};

\draw[flow] (c-alice.east) -- (c-bob-nc.west);
\draw[flow] (c-alice.east) -- (c-bob-c-direct.west);

\node[endpoint] (c-charlie-c3) at (6.6,-2.75) {
    Charlie\\
    \textit{Custodial}
};

\node[state] (c-charlie-nc4) at (6.6,-4.05) {
    Charlie\\
    \textit{Self-custodial}
};

\node[endpoint] (c-charlie-c4) at (9.9,-4.05) {
    Charlie\\
    \textit{Custodial}
};

\node[state] (c-bob-c) at (6.6,-5.45) {
    Bob\\
    \textit{Custodial}
};

\draw[flow] (c-bob-nc.east) -- (c-charlie-c3.west);
\draw[flow] (c-bob-nc.east) -- (c-charlie-nc4.west);
\draw[flow] (c-charlie-nc4.east) -- (c-charlie-c4.west);
\draw[flow] (c-bob-nc.east) -- (c-bob-c.west);

\node[case] at (8.15,-2.75) {UC-3};
\node[case] at (11.45,-4.05) {UC-4};

\node[endpoint] (c-charlie-c5) at (9.9,-5.0) {
    Charlie\\
    \textit{Custodial}
};

\node[state] (c-charlie-nc6) at (9.9,-6.25) {
    Charlie\\
    \textit{Self-custodial}
};

\node[endpoint] (c-charlie-c6) at (13.2,-6.25) {
    Charlie\\
    \textit{Custodial}
};

\draw[flow] (c-bob-c.east) -- (c-charlie-c5.west);
\draw[flow] (c-bob-c.east) -- (c-charlie-nc6.west);
\draw[flow] (c-charlie-nc6.east) -- (c-charlie-c6.west);

\node[case] at (11.45,-5.0) {UC-5};
\node[case] at (14.75,-6.25) {UC-6};

\node[endpoint] (c-charlie-c7) at (6.6,-7.1) {
    Charlie\\
    \textit{Custodial}
};

\node[state] (c-charlie-nc8) at (6.6,-8.4) {
    Charlie\\
    \textit{Self-custodial}
};

\node[endpoint] (c-charlie-c8) at (9.9,-8.4) {
    Charlie\\
    \textit{Custodial}
};

\draw[flow] (c-bob-c-direct.east) -- (c-charlie-c7.west);
\draw[flow] (c-bob-c-direct.east) -- (c-charlie-nc8.west);
\draw[flow] (c-charlie-nc8.east) -- (c-charlie-c8.west);

\node[case] at (8.15,-7.1) {UC-7};
\node[case] at (11.45,-8.4) {UC-8};

% ================================================================
% CROSS-BORDER TRANSFER PATHS
% ================================================================

\node[sectiontitle] at (0,-10.05) {Cross-border transfer paths};

\node[state] (x-alice) at (0,-11.65) {
    Alice\\
    \textit{Self-custodial}
};

\node[state] (x-exchange) at (4.2,-11.65) {
    Exchange service\\
    \textit{Custodial}
};

\node[endpoint] (x-bob-c9) at (8.4,-11.0) {
    Bob\\
    \textit{Custodial}
};

\node[state] (x-bob-nc10) at (8.4,-12.3) {
    Bob\\
    \textit{Self-custodial}
};

\node[endpoint] (x-bob-c10) at (12.6,-12.3) {
    Bob\\
    \textit{Custodial}
};

\draw[flow] (x-alice.east) -- (x-exchange.west);
\draw[flow] (x-exchange.east) -- (x-bob-c9.west);
\draw[flow] (x-exchange.east) -- (x-bob-nc10.west);
\draw[flow] (x-bob-nc10.east) -- (x-bob-c10.west);

\node[case] at (9.95,-11.0) {UC-9};
\node[case] at (14.15,-12.3) {UC-10};

% ================================================================
% ADVERSARIAL SCENARIO
% ================================================================

\node[sectiontitle] at (0,-14.15) {Adversarial scenario};

\node[state, text width=2.8cm, minimum height=1.05cm]
    (t-alice) at (0,-16.4) {
    Alice\\
    \textit{Self-custodial}\\
    \textbf{Key compromised}
};

\node[state, text width=2.8cm, minimum height=1.05cm]
    (t-adversary) at (5.2,-16.4) {
    Adversary\\
    \textit{Self-custodial}
};

\node[state, text width=3.0cm, minimum height=1.05cm]
    (t-intake) at (10.4,-16.4) {
    Custodial institution\\
    \textit{Evaluates intake request}
};

\node[endpoint, text width=2.7cm, minimum height=1.0cm]
    (t-accepted) at (15.2,-15.1) {
    Intake accepted
};

\node[endpoint, text width=2.7cm, minimum height=1.0cm]
    (t-held) at (15.2,-17.7) {
    Intake restricted\\
    \textit{Pending review}
};

\draw[flow] (t-alice.east) -- (t-adversary.west);
\node[edgeannotation] at (2.6,-15.35) {
    Unauthorized update using\\
    the compromised controller key
};

\draw[flow] (t-adversary.east) -- (t-intake.west);
\node[edgeannotation] at (7.8,-15.35) {
    Presents asset for\\
    redemption or intake
};

\draw[flow]
    (t-intake.east)
    to[out=23,in=180]
    (t-accepted.west);
\node[branchannotation] at (12.9,-15.55) {
    No active\\
    theft report
};

\draw[flow]
    (t-intake.east)
    to[out=-23,in=180]
    (t-held.west);
\node[branchannotation] at (12.9,-17.2) {
    Active\\
    theft report
};

\node[case] at (16.85,-16.4) {UC-11};

\end{tikzpicture}
}% end resizebox
}% end makebox

\caption{Branching index of the transfer paths and adversarial scenario
considered in the use-case analysis. \textit{Self-custodial} denotes direct
possession and control of the relevant digital asset. \textit{Custodial} denotes
possession or control mediated by a custodial institution. Common initial
transfer stages are shown once and then divided into separate branches for use
cases that continue differently. The asset-theft scenario is represented
separately because key compromise is an adversarial condition that may affect
multiple transfer paths rather than a distinct custody configuration.}
\label{fig:use-case-index}

\end{figure}

Figure~\ref{fig:use-case-index} identifies the complete set of transfer paths
considered in this paper. It serves as an index of the relevant custody
configurations and shows the common initial stages shared by related use
cases. A self-custodial node represents direct possession and control of the
digital asset by the named participant. A custodial node represents value
held or controlled through a custodial institution on behalf of that
participant.

The subsections that follow develop only the most salient cases in detail, which avoids repeating substantially equivalent transfer mechanics.

% FIGURE STARTS

% ------------------------------------------------------------------
% Use-case branching index
% ------------------------------------------------------------------

% The H placement specifier prevents later text from moving ahead of
% the figure. If there is not enough room on the current page, the
% complete figure begins on the next page.

% FIGURE ENDS

The indexed paths fall into four groups. UC-1 and UC-2 describe the simplest
transitions from self-custodial control to a custodial endpoint. In UC-1, Alice
transfers value directly to Bob in custodial form. In UC-2, Bob first receives
the asset under self-custodial control and subsequently presents it for
custodial intake.

UC-3 to UC-8 describe longer domestic chains. UC-3 and UC-4 preserve
self-custodial circulation beyond the initial transfer before value reaches a
custodial endpoint. UC-5 and UC-6 introduce an intermediate transition into
Bob's custody, after which value is transferred onward either directly into
Charlie's custody or first into Charlie's self-custodial control. UC-7 and
UC-8 begin with direct custodial receipt by Bob and distinguish between an
onward custodial transfer and a subsequent return to self-custodial
circulation.

UC-9 and UC-10 extend the analysis to cross-border transfer paths involving a
cross-currency exchange service that sources liquidity on behalf of the sender
of an asset, by identifying a recipient of a countervailing transaction and
sending the asset to that recipient along with appropriate compliance data about
the exchange, such as the source currency (see section~\ref{ss:fx}).  UC-9
terminates directly in Bob's custodial environment.  UC-10 instead delivers the
exchanged value to Bob under self-custodial control before a later custodial
intake.

UC-11 differs from the preceding cases because it does not define an additional
custody path. Instead, it examines how the architecture behaves when an
adversary obtains the signature key controlling a self-custodial asset.  Because
possession of the controlling key may enable the adversary to produce an
otherwise structurally valid update, the transfer layer cannot generally
distinguish theft from an update authorized by the legitimate holder. The
scenario therefore focuses on later containment at custodial intake or
redemption, where applicable registry-backed theft reports or supervisory flags
may prevent immediate conversion and trigger further review.  Our approach
supports the option for institutions or custodians to impose a delay on incoming
assets where necessary, as a way to mitigate the risk of trafficking stolen
assets.

% Might want to describe the "error" case as well, wherein a user/merchant/payee
% receives an asset without a valid proof of compliance.  This genrally means
% that the recipient was not authorised and might need to get an affidavit or ex
% post certification or pay a penalty, etc.: ADDRESS IN THE FOLLOWING PARAGRAPH

The preceding cases assume that the recipient obtains the compliance evidence the
applicable policy requires. A recipient may nevertheless be handed an asset without a
valid proof of compliance, for example, because the recipient was not authorised for the
transaction and the CAS therefore withheld its attestation, or because the sender
committed to a transaction that did not match the recipient's commitment. Because the
integrity layer is oblivious to transaction contents, nothing at the transfer layer
prevents this; the deficiency surfaces when the recipient verifies the asset or presents
it for custodial intake or redemption, where it may be rejected, held for review, or
restricted (Section~\ref{sec:core-protocol-rules}). Whether and how the recipient can then
cure the deficiency, for example through an affidavit, an ex post certification, or a
penalty, depends on the applicable legal and institutional process rather than on the
protocol itself.

The ten transfer paths are not treated as ten independent protocols. They are
combinations of a smaller number of recurring operations: self-custodial
transfer, custodial intake, custodial transfer, return to self-custodial
circulation, and cross-border exchange. UC-11 is considered separately as an
adversarial scenario that may arise in connection with multiple transfer
paths. The detailed use cases below are selected according to whether they
introduce a distinct compliance, acceptance, or convertibility question.
Shorter or structurally equivalent paths are addressed through the
corresponding representative case.

\subsection{Representative Use Cases Selected for Detailed Analysis}
\label{sec:representative-use-cases}

Developing every path in the use-case index independently would introduce substantial repetition, because several cases differ only through the addition or omission of an intermediate custodial or self-custodial stage. The analysis therefore focuses on representative cases selected for introducing a distinct architectural, compliance, acceptance, or convertibility question.

The following cases are examined in detail:

\begin{itemize}
    \item \textbf{Direct self-custodial payment to a custodial recipient
    (UC-1).} This case establishes the baseline interaction between a
    self-custodial payer and a recipient operating through a custodial
    institution. It is used to explain the complete transfer sequence,
    including construction of the proposed asset update, compliance
    evaluation, controller authorization, integrity processing, and
    custodial acceptance.

    \item \textbf{Multi-hop self-custodial circulation followed by custodial
    intake (UC-4).} This case illustrates how an asset may pass through
    successive self-custodial holders before being presented to a custodial
    institution. It is used to examine continuity of provenance, accumulation
    and verification of compliance evidence, and the conditions governing
    later convertibility.

    \item \textbf{Cross-border exchange followed by self-custodial receipt
    and redemption (UC-10).} This case extends the architecture to a
    cross-border path involving an exchange service, delivery of value to a
    self-custodial recipient, and subsequent custodial intake. It is used to
    examine the interaction between exchange operations, jurisdictionally
    applicable compliance conditions, and redemption.

    \item \textbf{Asset theft and key compromise.} This adversarial case does
    not represent an additional custody configuration. Instead, it examines
    the consequences of compromise of the signature key controlling a
    self-custodial asset, the limits of transfer-layer detection, and the
    potential role of custodial intake, external reports, and regulatory
    registries in later containment.
\end{itemize}

UC-1 is presented with a detailed interaction flow because it establishes the basic sequence reused by the remaining transfer cases, which are described in text where their mechanics are substantially the same.

\subsection{Retail Payment from a Self-Custodial Wallet to a Custodial Recipient (UC-1)}
\label{sec:uc-1}

\subsubsection{Scenario}

Alice controls an asset in state $A_{j-1}$ through a self-custodial wallet and uses it to make a retail payment at a point of sale. The merchant, Bob, receives payment as credit to an account provided by his custodial institution $B_b$, where $B_b$ denotes Bob's bank. Alice communicates with the merchant's point-of-sale terminal rather than directly with the institution, the Compliance Attestation Service, or the integrity provider.

The custodial institution provides the POS with the next controller verification key
\(k_{j+1}\). This key may be requested when the transaction is initiated or
drawn from a set of verification keys provisioned to the POS in advance.

% Retail-payment interaction sequence

\begin{figure}[H]
\centering

\makebox[\textwidth][c]{
\resizebox{1.06\textwidth}{!}{
\begin{tikzpicture}[
    actor/.style={
        draw,
        rounded corners=2pt,
        minimum width=2.8cm,
        minimum height=0.75cm,
        align=center,
        font=\small
    },
    lifeline/.style={
        densely dashed,
        gray
    },
    message/.style={
        -{Latex[length=2mm]},
        thick
    },
    response/.style={
        -{Latex[length=2mm]},
        thick,
        dashed
    },
    msg/.style={
        font=\scriptsize,
        align=center,
        fill=white,
        inner sep=1.5pt
    },
    action/.style={
        draw,
        rounded corners=2pt,
        fill=gray!8,
        font=\scriptsize,
        align=center,
        text width=2.8cm,
        inner sep=3pt
    }
]

\node[actor] (alice) at (0,0) {
    Alice's wallet
};

\node[actor] (pos) at (4.1,0) {
    Merchant POS
};

\node[actor] (cas) at (8.2,0) {
    Compliance\\
    Attestation Service
};

\node[actor] (integrity) at (12.3,0) {
    Integrity provider
};

\node[actor] (bank) at (16.4,0) {
    Merchant's custodial\\
    institution
};

\draw[lifeline] (alice.south) -- ++(0,-18.1);
\draw[lifeline] (pos.south) -- ++(0,-18.1);
\draw[lifeline] (cas.south) -- ++(0,-18.1);
\draw[lifeline] (integrity.south) -- ++(0,-18.1);
\draw[lifeline] (bank.south) -- ++(0,-18.1);

\draw[response]
    (16.4,-1.5)
    --
    node[msg, above, text width=3.4cm]
    {
        Hash of next controller attachment point and public key
        \(h(G_{L,i},k_{j+1})\)
    }
    (4.1,-1.5);

\draw[message]
    (4.1,-2.5)
    --
    node[msg, above, text width=3.7cm]
    {
        Submit \(\tilde{F}_j\) for
        compliance evaluation
    }
    (8.2,-2.5);

\node[action] at (8.2,-3.5) {
    Determine applicable
    predicates and perform
    required checks
};

\draw[response]
    (8.2,-5.0)
    --
    node[msg, above, text width=3.7cm]
    {
        Compliance attestation
        \(\alpha_j\), or rejection
    }
    (4.1,-5.0);

\node[action] at (4.1,-6) {
    Embed \(\alpha_j\) in \(u_j\)
};

\draw[message]
    (4.1,-7.5)
    --
    node[msg, above, text width=3.5cm]
    {
        Proposed update including compliance attestation, anchor, and key \((F_j)\)
    }
    (0,-7.5);

\node[action] at (0,-9) {
    Construct proposed
    asset update vector \(U_j\) by attaching \(u^s_j\) and \(h(A_{j-1})\)
};

\draw[message]
    (0,-11)
    --
    node[msg, above, text width=3.5cm]
    {
        Signed asset update
        vector \(U_j\)
    }
    (4.1,-11);

\draw[message]
    (4.1,-12)
    --
    node[msg, above, text width=3.7cm]
    {
        Submit signed update
        for integrity processing
    }
    (12.3,-12);

\draw[response]
    (12.3,-13)
    --
    node[msg, above, text width=3.3cm]
    {
        Integrity evidence
    }
    (4.1,-13);

\draw[response]
    (4.1,-14)
    --
    node[msg, above, text width=3.2cm]
    {
        Forward integrity
        evidence
    }
    (0,-14);

\draw[message]
    (4.1,-15)
    --
    node[msg, above, text width=4.0cm]
    {
        Submit updated asset
        for custodial intake
    }
    (16.4,-15);

\draw[response]
    (16.4,-16)
    --
    node[msg, above, text width=3.7cm]
    {
        Asset accepted, rejected,
        or held for review
    }
    (4.1,-16);

\draw[response]
    (4.1,-17)
    --
    node[msg, above, text width=3.1cm]
    {
        Payment confirmed
        or declined
    }
    (0,-17);

\end{tikzpicture}
}
}

\caption{Retail payment from Alice's self-custodial wallet to a merchant
receiving the asset through a custodial institution. The point-of-sale
terminal mediates the compliance, integrity, and custodial-intake
interactions.}
\label{fig:uc1-sequence}

\end{figure}
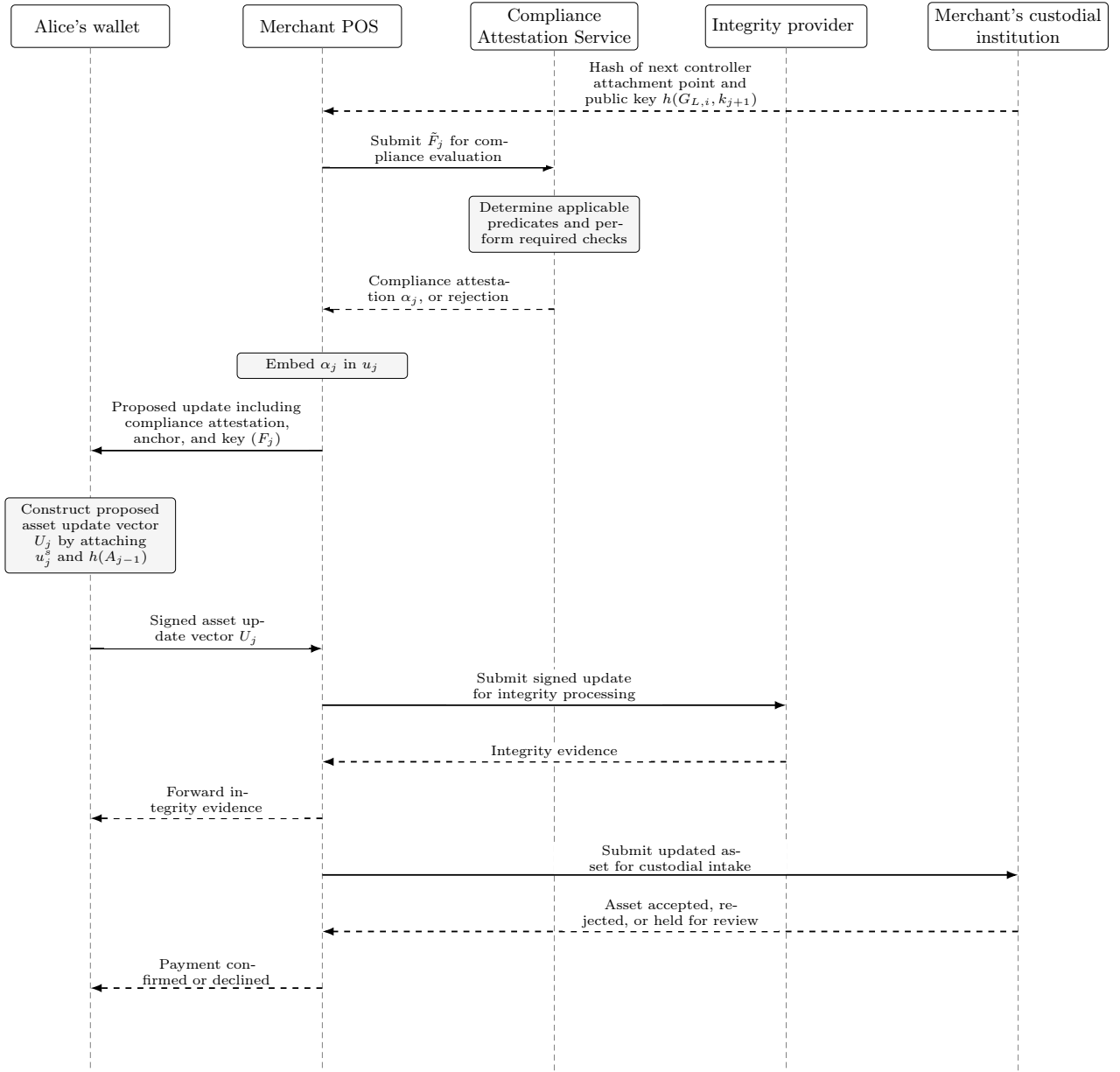

\subsubsection{Event Trace}

To create a proposed asset update with a compliance attestation, the point-of-sale (POS)
system first creates $\tilde{F}_j$, which is a commitment to the proposed asset update.
The proposed asset update includes a preliminary version of the transaction context
\(\tilde{u}_j\), the one-time key of the receiver \(k_{j+1}\), and the anchor point
\(G_{j,0}\):

\begin{equation}
\tilde{F_j} \leftarrow \tilde{u}_j\,||\,h(G_{j,0},k_{j+1})
\end{equation}

The merchant’s authorisation credential identifies the merchant and may specify
constraints relevant to the transaction. These constraints may include permitted
transaction categories, value or volume limits, jurisdictional conditions, and the period
during which the authorisation remains valid.

The POS submits $\tilde{F}_j$ to the Compliance Attestation Service \(\mathcal{S}\), which
determines the applicable predicate set

\begin{equation}
C_j \leftarrow \{r_1,\ldots,r_m\}
\end{equation}

and evaluates the proposed transfer. The request carries merchant-side context only: the
merchant's authorisation credential, the amount and category of the sale, and the
commitment $\tilde{F}_j$ to the proposed update. Alice has not yet taken part, so the
service learns nothing about her. If the proposed transfer is out of line with the
merchant's authorisation or recent activity, for example, because it is unusually large,
or because the recipient has approved too many transfers in a short period of time, then
the CAS can withhold the attestation or ask for further evidence that the sale falls
within the merchant's authorised categories and limits. It could also ask for Alice's
identity or other details, but such a request would reach her only through the POS, and
she is free to decline. If the attestation is withheld, then the payer can refuse to
authorise the transaction.

Some checks may be performed using information already embedded in \(\tilde{u}_j\),
including validation of the merchant's authorisation credential and comparison of the
transaction amount with an applicable limit. Other checks may require current external
information, including credential revocation, sanctions status, cumulative transaction
thresholds, merchant activity, or an active theft report.

For example, the POS may include evidence that the payment is transaction 26 of 2,000
processed by the merchant during the relevant day. Such evidence may support a threshold
check where the counter is produced through a trusted or verifiable mechanism.

If the applicable predicates are satisfied, \(\mathcal{S}\) issues the compliance
attestation \(\alpha_j\), which denotes the compliance attestation for update \(j\): a
signed statement confirming that the proposed asset update has satisfied the applicable
compliance rules under the relevant policy or registry version.  Following the
update-binding rules described in Section~\ref{sec:compliance-attestation-binding}, the
attestation is associated with the relevant transaction facts, the ledger reference
\(G_{j,0}\), and the next controller verification key \(k_{j+1}\).  The compliance attestation
attests that the applicable predicates were satisfied.  If the compliance evaluation
fails, then no attestation is issued.

The Compliance Attestation Service returns the compliance attestation \(\alpha_j\) to the
POS, which creates the finalised transaction context \(u_j\) by combining the preliminary
version of the transaction context \(\tilde{u_j}\) with the compliance attestation
\(\alpha_j\):

\begin{equation}
u_j \leftarrow h(\tilde{u_j}) \,||\, \alpha_j
\end{equation}
\begin{equation}
F_j \leftarrow u_j\,||\,h(G_{j,0},k_{j+1})
\end{equation}

The POS then forwards \(h(F_j)\) to Alice’s wallet.  If Alice's wallet requests the
compliance attestation, the POS may also provide \(u_j\), which includes the hash of the
preliminary version of the transaction context \(h(\tilde{u}_j)\) and the attestation
\(\alpha_j\), along with the hash \(h(G_{j,0},k_{j+1})\) needed to validate that it
matches \(h(F_j)\).  The preliminary version of the transaction context \(\tilde{u}_j\)
may itself also include information that Alice requires the merchant to provide
specifically, such as the applicable jurisdiction or the merchant’s authorisation
credential.  In this case, the POS system can provide that information separately to
Alice's wallet, along with proof that such information is covered by the hash
\(h(\tilde{u_j})\).

Alice's wallet then combines \(h(F_j)\) with its own transaction context \(u^s_j\) and the
previous state of the asset \(A_{j-1}\) and constructs the update vector \(U_j\) (see
Equation~\ref{e:U_j}).

Alice's controller signature contained within \(U_j\) authorises the transfer of control.
Alice's wallet sends the previous asset state \(A_{j-1}\) along with the signed asset
update vector \(U_j\) to the POS. The POS then submits the following items to the
integrity provider identified in \(A_{j-1}\):

\begin{itemize}

\item the verification key \(k_j\),

\item the hash of the signed update \(h(u^s_j\,||\,h(F_j)\,||\,h(A_{j-1}))\), and

\item the signature on that hash made by the signature key corresponding to \(k_j\).

\end{itemize}

The integrity provider processes the update and returns the resulting integrity evidence
to the POS. The POS forwards that evidence to Alice's wallet, allowing the wallet to
verify that the update has been registered.

\subsubsection{Custodial Intake}

After the integrity evidence has been obtained, the POS submits the updated
asset to \(B_b\) for custodial intake. 

If the intake conditions are satisfied, \(B_b\) accepts control of the asset
on behalf of the merchant and informs the POS that the payment has been
accepted. The POS then confirms completion of the payment to Alice.

If the asset is rejected or held pending review, the POS informs Alice that
the payment has not been completed.

\subsubsection{Protocol Effect}

The POS acts as the merchant-side intermediary for the payment. It provides
the transaction context to Alice, forwards the proposed asset update vector
for compliance evaluation, returns the resulting attestation to the wallet,
submits the signed update for integrity processing, and presents the updated
asset to the merchant's custodial institution.

Alice's wallet remains responsible for finalising and authorising the asset
update. The Compliance Attestation Service evaluates the applicable compliance
conditions, the integrity provider processes the signed update hash, and the
custodial institution determines whether to accept control of the asset on
behalf of the merchant.

\subsection{Multi-hop self-custodial circulation followed by custodial intake (UC-4)}

This case illustrates how an asset may circulate through successive self-custodial holders before being presented to a custodial institution for intake. At each hop, the new prospective controller provides a hash comprehending a transaction context \(u_j\), anchor \(G_{j,0}\), and verification key \(k_j\); the current controller adds this hash to its transaction context \(u^s_j\) and the previous state of the asset \(A_{j-1}\), obtains a compliance attestation \(\alpha_j\) where the applicable predicates are satisfied, signs the completed update, and submits it for integrity processing.

Custodial intake does not occur immediately after the first transfer: the asset remains in self-custodial circulation across multiple state updates. The system must therefore preserve a verifiable basis for later convertibility, even though no custodial institution takes control of the asset during the intermediate transfers. For this reason, the compliance evaluation may require a redeemer authorisation signature, which designates the custodial institution authorised to redeem or take in the asset at a later stage, for example through the identifier of a regulated financial institution recorded in an authoritative registry. It does not transfer control to that institution at the time of the self-custodial payment; it establishes the institutional endpoint whose approval is required for the asset later to be refreshed or converted into account-based or institutionally mediated monetary form. Once introduced, the constraint becomes part of the asset's compliance-relevant state: subsequent self-custodial updates may continue to transfer control, but must preserve it, and an update that removes, replaces, or contradicts it fails the applicable compliance evaluation.

Each intermediate update thus extends the provenance chain while accumulating compliance evidence. The integrity provider verifies the structural continuity of the asset state, and the embedded attestations show that each transfer satisfied the applicable rule set. When the asset is presented to the designated custodial institution, that institution can evaluate its authenticity, the continuity of the provenance chain, the validity of the accumulated attestations, and consistency with the redeemer authorisation signature. This use case therefore separates transfer admissibility from later convertibility: the asset may keep circulating provided each update satisfies the applicable predicates, while refreshment or conversion remains governed by the redeemer authorisation constraints and the institution's intake checks at the point of presentation.

\subsection{Cross-border exchange followed by self-custodial receipt and redemption (UC-10).} 
\label{ss:fx}

Like cash, money within the USO asset system cannot functionally cross borders, because the currency of one jurisdiction will not be accepted for redemption by the financial institutions of a different jurisdiction. Transferring money from an economy that uses currency \(X_a\) to one that uses a different currency \(X_b\) therefore requires an exchange mechanism, even if one currency is pegged to the other. Exchange services can facilitate such transfers by matching desired transfers from \(X_a\) to \(X_b\) with countervailing transfers from \(X_b\) to \(X_a\); these depend upon the presence of liquidity and cannot be instantaneous for both parties.

The process of transferring a token with quantity \(q_a\) from a payer \(P_a\) in jurisdiction \(a\) with currency \(c_a\) to a recipient \(R_b\) in jurisdiction \(b\) with currency \(c_b\) has three stages. It is assumed that the payer has satisfied the relevant compliance requirements in the jurisdiction using currency \(c_a\). In the first stage, payer \(P_a\) transfers the token to an exchange \(X\), whose operator operates in both jurisdiction \(a\) and jurisdiction \(b\). In the second stage, exchange \(X\) matches the desired quantity \(q_a\) against a matching quantity \(q_b\) of a desired countervailing transaction from a payer \(P_b\) in jurisdiction \(b\) with currency \(c_b\) seeking to pay a recipient \(R_a\) in jurisdiction \(a\) with currency \(c_a\). Without loss of generality, the countervailing transaction can be substituted by multiple transactions, possibly from different payers, that sum to \(q_b\). The exchange operator \(X\) may capture a fraction of the spread in the exchange between \(c_a\) and \(c_b\).

The token from payer \(P_a\) with quantity \(q_a\) is then transferred from the exchange \(X\) to recipient \(R_a\), and the token from payer \(P_b\) with quantity \(q_b\) is transferred from the exchange \(X\) to recipient \(R_b\). In the present use case, the relevant recipient receives the asset in a self-custodial wallet before any later redemption or custodial intake occurs. The exchange \(X\) collects whatever compliance data may be required from both payers without necessarily requiring them to provide their identities, and the identity of the exchange is recorded in the provenance of both tokens.

The exchange service therefore contributes compliance-relevant evidence to the transfer path. This may include exchange-related authorisation credentials, jurisdictional metadata, transaction classification, and evidence that the exchange operation is permitted under the relevant regulatory regimes. Such information may be included in the transaction context embedded in \(u_j\), together with the next controller verification key and any applicable redemption-related constraint.

Without loss of generality, the use case may be illustrated through a UK--US transfer path. In that case, the derived constraint set may be represented as:

\begin{equation}
C_j = C_j^{(\mathsf{UK})} \cup C_j^{(\mathsf{US})}
\end{equation}

where \(C_j^{(\mathsf{UK})}\) denotes the constraints associated with the UK-side jurisdictional attachment, and \(C_j^{(\mathsf{US})}\) denotes the constraints associated with the US-side jurisdictional attachment or designated redemption endpoint.

The mechanics of the asset update remain structurally similar to the domestic self-custodial transfer cases. What changes is the applicable constraint set, which now reflects the regulatory regimes governing the legally relevant entities, including the payer's jurisdiction and that of the designated custodial institution. The compliance predicate is evaluated against the aggregated rule set, admission of the update requires simultaneous satisfaction of all applicable constraints, and the resulting attestation \(\alpha_j\) certifies that the transfer is admissible under each implicated jurisdiction.

If the asset is later submitted to the designated US custodial institution for credit into account-based form, convertibility conditions are evaluated in light of US supervisory requirements, while preserving verification of the full compliance provenance chain accumulated during prior state updates. Redemption occurs only after the self-custodial recipient presents the asset to the designated custodial endpoint. Cross-border operation thus does not alter the mechanics of transfer; it expands the constraint set from which admissibility is derived, while regulatory observability and convertibility remain governed by the designated custodial endpoint, whose jurisdiction determines the supervisory interface at institutional intake. Where required by the audit-log operational model, the convertibility certificate \(\pi(A_j)\) may be committed to an append-only log accessible to the supervisory authority governing the designated institution, and inter-jurisdictional information sharing occurs through established regulatory channels.

\subsection{Asset theft and key compromise (UC-11).}

This adversarial case examines the consequences of compromise of the signature key controlling a self-custodial asset, and the limits of transfer-layer detection.

Suppose an adversary \(u_x\) obtains the signature key controlling asset state \(A_{j-1}\), previously held by victim \(u_i\). From the protocol's perspective, possession of the controlling signature key is sufficient to construct a syntactically valid ownership update. The resulting transfer request may therefore be indistinguishable from an authorised transfer at the level of structural verification.

The adversary constructs an update vector \(U_j\), designating a new controlling verification key and including the required transaction descriptors. The applicable constraint set

\begin{equation}
C_j = \{r_1,\ldots,r_m\}
\end{equation}

is derived in the ordinary manner from registry-backed contextual information, and the compliance predicate \(C(A_{j-1},F_j)\) is evaluated. If the predicate is satisfied, the Compliance Attestation Service issues the compliance attestation \(\alpha_j\), and the signed update may be submitted for integrity processing. Upon successful validation of provenance continuity, successor state \(A_j\) is recognised.

Accordingly, theft cannot generally be prevented at the transfer layer once the controlling signature key has been compromised. The transfer layer enforces protocol-defined predicates and structural validity; it does not determine whether the person exercising the key is acting with legitimate authority.

Containment may instead arise later, if the adversary attempts to submit \(A_j\) for custodial intake or redemption at a designated institution \(B_j\). At that point, the institution may verify the authenticity of the asset, the continuity of the provenance chain, the validity of accumulated compliance attestations, and consistency with any embedded redeemer authorisation constraint.

In addition to those structural checks, the convertibility evaluation may incorporate external reports or registry-based flags. For example, a prior controller or supervisory authority may report the asset as stolen, and that report may be associated with the asset identifier or provenance chain within the compliance infrastructure. If the asset is subject to an active theft flag, convertibility need not succeed pending investigation. The institution may deny immediate conversion or place the asset into temporary institutional custody while supervisory review is conducted.

This case therefore illustrates the separation between transfer-layer admissibility and convertibility-layer containment. A key-compromise event may produce a structurally valid transfer, but later custodial intake creates a verifiable institutional decision point at which external reports, regulatory registries, and supervisory obligations can support containment, logging, or temporary immobilisation.

\section{Discussion and Implications}
\label{sec:discussion}

\subsection{From Custodial Visibility to Verifiable Compliance}

The architecture reframes the relationship between privacy, custody, and regulatory compliance. In conventional systems, compliance is tied to custodial visibility: the institution that holds user funds also observes transactions, applies screening rules, keeps account records, and supplies information to supervisors. This is familiar but costly and trust-intensive. Users must trust custodians with sensitive data, and regulators must rely on records produced by those same institutions, often with limited independent means of verifying them. Privacy and auditability therefore both depend on institutional conduct rather than on system design.

\begin{center}
\fbox{
\begin{minipage}{0.92\linewidth}
\textbf{Central architectural implication.}
The conventional coupling between custodial control, transaction visibility, and regulatory auditability is not technically necessary. Compliance-relevant conditions can be evaluated at the level of asset state transitions, and the resulting evidence can be embedded directly into the asset state.
\end{minipage}
}
\end{center}

This builds on the USO model, in which assets are held by users rather than by the ledger, notarisation services remain oblivious to transaction contents, and a proof of provenance shows that updates were incorporated into an authoritative history \cite{goodell_toliver_nakib_scalable_payments_2021,lynham_sark_2025}. Here the asset also carries verifiable compliance evidence as it evolves, while the integrity layer stays concerned only with the structural validity and commitment of updates: it need not inspect transaction contents, identify counterparties, or keep records of payment histories.

The relevant distinction is between compliance based on institutional visibility over payment flows and compliance based on verifiable evidence attached to asset state transitions. Protocol-level compliance could be implemented as centralised transaction surveillance, but it need not be. The architecture still relies on authoritative regulatory bodies, recognised registries, and designated attestation services. However, it does not require the ledger, issuer, payment processor, or any other party to accumulate a comprehensive view of transaction activity. Their role is to define, certify, or verify the conditions under which asset state transitions are compliant, not to process every payment as a visible account-level transaction. Verification rests on cryptographic evidence and registry-backed rules, not on general access to payment flows.

\subsection{Institutional Allocation of Functions}

If compliance is not located in general transaction visibility, the institutional question becomes where legal authority, monetary convertibility, and compliance verification should sit. The architecture places them at the edges of the transaction. Financial institutions operate at the boundary between conventional custodial deposits and the self-custodial domain, performing regulated withdrawal and redemption and applying admissibility checks on conversion. Regulatory authorities define the predicate sets, supervise the registries, and set the audit configuration. Compliance Attestation Services act as designated evaluators, issuing signed evidence that a proposed asset state transition satisfies the applicable rules.

The resulting structure separates three functions that are ordinarily bundled together:

\begin{itemize}
    \item\textbf{Custody.}
    Custody concerns who controls the asset.

    \item\textbf{Compliance evaluation.}
    Compliance evaluation concerns whether the evidence attached to a proposed update demonstrates satisfaction of the applicable predicates.

    \item\textbf{Payment execution.}
    Payment execution concerns whether the asset state is validly updated and committed.
\end{itemize}

Separating these functions enables a non-custodial architecture while preserving identifiable points of regulatory control. Compliance can be required at transfer, at redemption, or at both points, depending on the adopted policy configuration.

\paragraph{Integrity provider and asset-carried evidence.} The payment processor or integrity provider supports only the validity and provenance of asset updates: it confirms that an update was validly committed and incorporated into the proof-of-provenance chain, without inspecting transaction contents, identifying counterparties, or evaluating compliance predicates. Compliance evidence is produced outside the integrity layer and embedded in the update context. In account-based systems such evidence usually sits outside the payment object, in institutional records and audit trails; here selected evidence becomes part of the asset's own state history, as structured attestations, registry references, and commitments sufficient for later verification by authorised parties, without disclosing all transaction data publicly. Auditability is therefore achieved through controlled disclosure and verifiability, not through comprehensive visibility.

\subsection{Enforcement and Institutional Implications}

Because compliance evidence can travel with the asset, enforcement can be configured in more than one way:

\begin{itemize}
    \item\textbf{Strict transfer-level configuration.}
    A proposed update is not accepted under the relevant policy unless the applicable compliance attestation is present.

    \item\textbf{Redemption-gated configuration.}
    An update may circulate within the self-custodial domain, while later convertibility into custodial form depends on the presence and validity of accumulated compliance evidence.
\end{itemize}

This second configuration is especially important because it preserves the possibility of oblivious transfer processing while using custodial intake as a later enforcement point. Assets that fail convertibility checks may be rejected, held for review, or otherwise restricted according to the applicable legal and institutional process.

\paragraph{Implications for financial institutions.}

For financial institutions, these configurations change the locus of involvement. Rather than being displaced, they remain responsible for admissibility and compliance checks at regulated points of entry, exit, and conversion; their role shifts from continuous mediation of every payment to regulated intake, redemption, issuance support, and risk review. They still provide the interface through which users acquire and redeem monetary value, but need not observe every later transfer for compliance to be verifiable.

\paragraph{Implications for regulatory design.}

For regulatory design, the architecture supports more modular rule implementation. Because rules live in versioned registries and signed attestations, regulatory change can occur through registry updates rather than changes to the asset protocol, while predicates, reference datasets, authorised attesters, and audit requirements evolve. This matters in regulated environments, where legal requirements change more often than foundational payment infrastructure.

\subsection{Areas Requiring Further Development}

The architecture nevertheless requires further development before it can be treated as a complete system specification. Four areas are especially important:

\begin{enumerate}[label=\textbf{\arabic*.},leftmargin=*]

    \item \textbf{Governance model.}

    The governance model must be defined. This includes the allocation of responsibilities among regulatory authorities, financial institutions, attestation services, registry operators, and audit-log administrators.

    \item \textbf{Operational specification.}

    The operational specification must be completed, including the structure of compliance envelopes, the procedures for signing and verifying attestations, and the relationship between embedded evidence and external audit records.

    \item \textbf{Proof-of-concept implementation.}

    The architecture must be implemented and tested through a proof of concept capable of demonstrating the interaction between wallets, attestation services, registries, integrity providers, and custodial intake mechanisms.

    \item \textbf{Business models and institutional opportunities.}

    The business models and institutional opportunities associated with this rearrangement of functions must be clarified, including the roles, incentives, liabilities, and service models that could support deployment.

\end{enumerate}

These open areas do not weaken the central claim. The contribution is not that institutions, governance, or legal process become unnecessary, but that those functions can be reorganised so that regulatory compliance does not depend on general custodial surveillance of payment activity.

\section{Conclusions and Future Work}
\label{sec:conclusion}

This paper has presented a reference architecture and core protocol rules for privacy-preserving, non-custodial digital payments designed to provide strong compliance assurances. The architecture builds on the USO asset model developed in prior work, in which digital assets are portable, stateful objects that carry their own proof of provenance and may be held directly by users. The contribution of this paper is not any one element, whether the USO asset model, signed attestations or embedded compliance, but their combination in a compliance architecture that does not assume that assets or transactions are recorded on a ledger. Compliance evidence is embedded into asset state transitions through signed attestations issued by authorised compliance services, while regulatory registries define the applicable predicates and reference data used in evaluation.

The model addresses a central tension in digital payment design. Existing electronic payment systems achieve auditability through custodial visibility, while cash-like systems preserve privacy but provide limited native compliance evidence. The architecture developed here offers a different approach: compliance conditions are evaluated and evidenced at the level of asset state transitions, without requiring a ledger, issuer, or payment processor to keep a comprehensive record of user transactions. Verification can then rely on cryptographic evidence, registry-backed rules, and accumulated attestations rather than on institutional visibility over payment activity.

The architecture rearranges institutional functions. Financial institutions remain responsible for withdrawal, redemption, custodial intake, and admissibility checks at regulated points of entry, exit, and conversion; regulatory authorities define and supervise compliance conditions; Compliance Attestation Services evaluate proposed updates and issue signed evidence; registry operators publish rule sets, reference data, authorised attesters, and versions; and integrity providers commit update records and support the proof-of-provenance chain while remaining oblivious to transaction contents.  This allocation lets privacy, self-custody, and auditability coexist within one protocol framework, and concentrates compliance work at defined lifecycle events, including attestation, intake, redemption, and conversion, because achieving non-repudiability of transfers does not require third parties to handle sensitive data, such as specific transactions or their contents.

The use cases show how the architecture applies across several transfer paths: a self-custodial wallet paying a custodial recipient, circulation through several self-custodial holders before intake, cross-border exchange with additional jurisdictional predicates, and movement between custodial and self-custodial environments. The relevant question is not only whether a transfer can occur, but under what conditions the resulting asset state remains admissible for later conversion, redemption, or institutional acceptance.

\subsection{Future Work}

Future work should proceed in four main directions:

\begin{enumerate}[label=\textbf{\arabic*.},leftmargin=*]

    \item \textbf{Governance.}

    A complete implementation requires an institutional model specifying who may define compliance predicates, operate registries, issue attestations, and maintain audit records, and how these roles are authorised, supervised, and updated. Embedded compliance depends not only on cryptographic verification but also on the institutional legitimacy of the rules, registries, and attesters behind it.

    \item \textbf{Operational specification.}

    The architecture requires a precise specification of the compliance envelope (its structure, required and optional fields, registry references, attestation format, and verification procedures) and a clearer account of how audit records are produced and inspected, how redemption and admissibility checks are performed, and how embedded evidence interacts with evidence held by external systems.

    \item \textbf{Proof-of-concept implementation.}

    A minimal prototype should include a wallet that constructs asset updates, a Compliance Attestation Service that evaluates registry-backed predicates, a registry layer, an integrity provider that commits updates and returns provenance evidence, and a custodial intake or redemption service that verifies accumulated compliance evidence. It would test the operational assumptions of the architecture and identify the minimum information needed to support both privacy and auditability. A formal security analysis of the properties argued informally in Section~\ref{sec:security-properties} should accompany it.

    \item \textbf{Business models and institutional opportunities.}

    The architecture does not remove the need for financial institutions, compliance services, registries, audit mechanisms, or wallet infrastructure; it assigns these functions differently across the payment lifecycle. Future work should examine the incentives, service models, liability structures, and operational responsibilities that could support deployment by financial institutions, attestation providers, registry operators, wallet providers, and audit-log administrators.

\end{enumerate}

The broader conclusion is that compliance need not be synonymous with custody, and privacy need not be synonymous with regulatory opacity. By embedding verifiable compliance evidence into self-custodial asset state, digital payment systems can preserve meaningful user privacy while remaining compatible with institutional oversight, regulatory change, and existing financial infrastructure.

\clearpage

\begingroup\small
\bibliography{references}
\endgroup

\end{document}